\documentclass{SciPost}

\hypersetup{
    colorlinks,
    linkcolor={red!50!black},
    citecolor={blue!50!black},
    urlcolor={blue!80!black}
}

\usepackage[bitstream-charter]{mathdesign}
\DeclareSymbolFont{usualmathcal}{OMS}{cmsy}{m}{n}
\DeclareSymbolFontAlphabet{\mathcal}{usualmathcal}

\fancypagestyle{SPstyle}{
\fancyhf{}
\lhead{\colorbox{scipostblue}{\bf \color{white} ~SciPost Physics Core }}
\rhead{{\bf \color{scipostdeepblue} ~Submission }}

\fancyfoot[C]{\textbf{\thepage}}
}

\usepackage{xcolor}
\usepackage{mathtools}
\usepackage{soul}
\newcommand{\NR}[1]{\textcolor{blue}{#1}}

\newcommand{\pa}[1]{\textcolor{black}{#1}}

\newcommand{\IR}{\mathbb{R}}

\begin{document}

\pagestyle{SPstyle}

\begin{center}{\Large \textbf{\color{scipostdeepblue}{
%%%%%%%%%% TODO: Write your article's title here
Diffusion of charged rods across 3D varying section channels\\
}}}\end{center}

\begin{center}\textbf{
% Use (full) first name (+ middle name initials) + surname format.
% Separate subsequent authors by a comma, omit comma and use "and" for the last author.
% Mark the corresponding author(s) with a superscript symbol in this order
% \star, \dagger, \ddagger, \circ, \S, \P, \parallel, ...
Nadja Ray\textsuperscript{1,2$\star$}, Alena Semkiv\textsuperscript{1}, 
Jens Harting\textsuperscript{3,4,5}, and Paolo Malgaretti\textsuperscript{3,5$\dagger$}
}\end{center}

\begin{center}
{\bf 1} Mathematical Institute for Machine Learning and Data Science, KU Eichstätt-Ingolstadt, Ingolstadt, Germany
\\

{\bf 2}  Department of Mathematics, Friedrich-Alexander-Universit\"at Erlangen-N\"urnberg, Cauerstr.\,11,
D-91058 Erlangen, 
Germany\\

{\bf 3} Helmholtz Institute Erlangen-N\"urnberg for Renewable Energy, Forschungszentrum J\"ulich, Cauerstr.\,1, 91058 Erlangen, Germany\\

{\bf 4} Department of Chemical and Biological Engineering, Friedrich-Alexander-Universit\"at Erlangen-N\"urnberg, Cauerstr.\,1,
D-91058 Erlangen, 
Germany\\

{\bf 5} Department of Physics, Friedrich-Alexander-Universit\"at Erlangen-N\"urnberg, Cauerstr.\,1,
D-91058 Erlangen, 
Germany\\

%%%%%%%%%% END TODO: AFFILIATIONS
% Provide email address of corresponding author(s)
%\\[\baselineskip]
$\star$ \href{mailto:email1}{\small nadja.ray@ku.de}\,,\quad
$\dagger$ \href{mailto:email2}{\small p.malgaretti@fz-juelich.de}
\end{center}

\section*{\color{scipostdeepblue}{Abstract}}
\textbf{\boldmath{%
%The fundamental understanding of particle transport through porous media is of high relevance in bio/biomedical, environmental and technical applications. Although the detailed shapes of the particles and surrounding pore space significantly influence their transport properties, they are still neglected in many investigations. Contrary, in this research, 
We analyze the transport of rod-like particles by diffusion and drift in a three-dimensional channel with varying circular or elliptic cross section. Applying the Fick-Jacobs approximation to the transport equation of the particles' probability distribution, we derive an effective one-dimensional substitute model and the associated free energy profile. Our results show that the data for the mean first passage time  of rods, once expressed as a function of the effective free energy barrier, collapse onto the same master curve as obtained for point or spherical particles.   
%\NR{I suggest to still either add an application as context in the beginning or conclude with some relevance of results, e.g. 
The observed universality provides a simple framework for predicting transport times of anisotropic particles in confined geometries without resolving the full multidimensional dynamics.
}}

\vspace{\baselineskip}

%%%%%%%%%% BLOCK: Copyright information
% This block will be filled during the proof stage, and finilized just before publication.
% It exists here only as a placeholder, and should not be modified by authors.
\noindent\textcolor{white!90!black}{%
\fbox{\parbox{0.975\linewidth}{%
\textcolor{white!40!black}{\begin{tabular}{lr}%
  \begin{minipage}{0.6\textwidth}%
    {\small Copyright attribution to authors. \newline
    This work is a submission to SciPost Physics Core. \newline
    License information to appear upon publication. \newline
    Publication information to appear upon publication.}
  \end{minipage} & \begin{minipage}{0.4\textwidth}
    {\small Received Date \newline Accepted Date \newline Published Date}%
  \end{minipage}
\end{tabular}}
}}
}
%%%%%%%%%% BLOCK: Copyright information

% For convenience during refereeing we turn on line numbers:
%\linenumbers

\vspace{10pt}
\noindent\rule{\textwidth}{1pt}
\tableofcontents
\noindent\rule{\textwidth}{1pt}
\vspace{10pt}
\section{Introduction}
\label{sec:introduction}

%\pa{add leonardo Dadgdug}

The transport of particles through porous materials plays a fundamental role in a wide range of natural and technological applications\cite{molnar2019colloid,tufenkji2007microbial}. Among other examples, ion transport through nanopores and porous membranes
is relevant to nanofluidic systems
\cite{schoch2008transport} and underpins desalination, water purification,
filtration, and selective separation processes
\cite{werber2016materials}. Likewise, ion transport across biological membranes
through channels, pumps, and transporters is essential for cellular
homeostasis and function
\cite{gouaux2005principles}.
In all these systems, macroscopic transport may depend sensitively on microscopic pore geometry, including constrictions, dead-end pores, and variations in the local cross section \cite{burada2009confined,bhattacharjee2019hopping,bordoloi2022vortices}. Direct observations of bacterial motion and solute dispersion in three-dimensional porous media have demonstrated that pore-scale trapping and geometrically induced flow structures can produce dynamics that cannot be inferred from the average porosity alone \cite{bhattacharjee2019hopping,bordoloi2022vortices}. Likewise, in~\cite{Ray2018,Prifling2023} heuristic porosity-diffusion laws were reviewed and numerical studies emphasize the limitation of using porosity as the primary descriptor of diffusive transport.
This geometric sensitivity is also exploited in analytical separation techniques like 
%size-based fractionation of colloidal particles and nanoparticles can also be achieved using 
hydrodynamic chromatography, size-exclusion chromatography, and field-flow fractionation, which separate suspended objects according to their hydrodynamic dimensions and, when coupled to suitable detectors, provide additional information about their morphology and internal structure
\cite{striegel2012hydrodynamic,pitkanen2016size,baalousha2011flow}.

What is more, non-spherical particles add an additional layer of complexity. In fact, direct pore-scale visualization has shown that particles with large aspect ratios, like microplastics, undergo interception, straining, rotation, and trapping, so their migration through soils, sediments, and granular filters can differ substantially from the transport of spherical microbeads~\cite{fouty2024microplastic}.  
Experiments have resolved anisotropic diffusion of ellipsoids and colloidal rods, quantified wall- and confinement-induced hydrodynamic hindrance, and shown that orientational dynamics modifies the longitudinal dispersion of elongated particles under flow \cite{han2006ellipsoid,mukhija2007dynamics,bitter2017confined,kumar2021taylor}. Accordingly, particles with the same volume or hydrodynamic size can exhibit markedly different mobility, retention, and passage probabilities. Shape-dependent transport has been observed in porous media and exploited in microfluidic separation, while studies of rod-like particles in macromolecular networks have revealed non-monotonic or unexpectedly rapid transport arising from the coupling between translation, rotation, and the confinement length scale \cite{masaeli2012shape,ma2020coupled,zhang2024sliding,xue2024coupling}. 
In corrugated channels, experiments on colloidal rods have further shown
that excluded-volume interactions generate an orientation-dependent entropic
free energy and promote alignment near bottlenecks, thereby modifying the
local diffusivity and mean first-passage time \cite{yang2019rods}.

%At the treatment-plant scale, measurements have found that conventional wastewater processes can remove a large fraction of the incoming microplastic load, but residual particles and fibres remain relevant because of the large effluent volumes continuously discharged \cite{carr2016fate,murphy2016wwtw,lares2018occurrence}. This has motivated detailed studies of shape- and size-selective segregation during treatment and of tertiary polishing technologies. Stepwise plant analyses and dedicated polishing studies have assessed biologically active filtration, disc filtration, rapid sand filtration, dissolved-air flotation, and membrane bioreactors for reducing the residual microplastic concentration in final effluents \cite{talvitie2017stepwise,talvitie2017solutions}. Recent slow-sand-filtration experiments using laundry wastewater further show that fibres' breakthrough depends strongly on filter-grain size, bed depth, and filtration rate \cite{gao2026slow}. 
%These applications provide an additional \NR{additional to what?}motivation for transport models that explicitly retain particle shape and orientational degrees of freedom.
Accordingly, even for noninteracting particles, a spatially varying channel produces an
effective free-energy landscape. In a constricted channel, the number of accessible transverse positions and orientations is reduced, giving rise to an entropic
barrier, in turn modulating the transport of particles through such channels. %\NR{Maybe add some motivation why effective models are crucial, but maybe we can find something better... 
However, numerical studies that directly respect particle size, shape, and orientation, see e.g \cite{Prignitz2014,Ruede2016,Meier2025}, are computationally demanding. The Fick--Jacobs approximation provides a systematic framework for
describing this effect by reducing the multidimensional Smoluchowski equation
to an effective one-dimensional transport equation along the channel axis
\cite{jacobs1967diffusion,zwanzig,Reguera,Malgaretti2013,dagdug2024book}. Systematic projection and
variational approaches have subsequently clarified the construction of the
reduced equation, the emergence of a position-dependent longitudinal
diffusivity, and the higher-order corrections generated by the channel
geometry
\cite{kalinay2005projection,kalinay2005variational,kalinay2006corrections,
dagdug2012projection,martens2011higher}. Comparisons with Brownian-dynamics
simulations have also delineated the regime, in which the modified
Fick--Jacobs description remains accurate
\cite{burada2007biased,berezhkovskii2015range,dagdug2024book}.
Within this reduced framework, confinement-controlled transport has been
characterized in terms of stationary currents, effective mobility,
permeability, diffusion resistance, and first-passage observables. Recent
work has related membrane permeability to the diffusion resistance of the
underlying pores
\cite{skvortsov2023permeability} and has investigated mean first-passage,
direct-transit, and looping times in conical and biconical channels with
entropic barriers or wells
\cite{perezespinosa2020,pompagarcia2022,
berezhkovskii2025biconical}. 
Reduced Fick--Jacobs descriptions have also been extended beyond point-like
tracers. They have been used to predict direction-dependent mean
first-passage times of charged tracers
\cite{malgaretti2016charged}, the translocation dynamics of polymers
\cite{bianco2016polymer}, and the permeability of varying-section channels
to neutral and charged rods \cite{Malgaretti_2020}. More recently,
the same framework has been used to determine splitting probabilities for
passive and active particles in three-dimensional corrugated channels
\cite{malgaretti2023splitting}. Nevertheless, the genuinely
three-dimensional transport of anisotropic rods, with two orientational
degrees of freedom and non-circular pore cross sections, remains
comparatively unexplored.

In the present work, we study the diffusion and force-driven transport of a thin rigid rod through a three-dimensional periodically modulated channel. The channel is described by either a circular or an elliptic cross section whose dimensions vary smoothly along its axis. We first derive the geometrical conditions that determine whether a given position and orientation of the rod are compatible with the channel boundaries. Starting from the translational and rotational Smoluchowski equation and assuming rapid local equilibration in the transverse and orientational degrees of freedom, we then derive an effective one-dimensional Fick--Jacobs equation. The reduced model contains a confinement-induced free energy, $A(x)$, and a local longitudinal diffusivity, $D(x)$, both obtained by averaging over the accessible positions and orientations of the rod.

We use this framework to investigate the equilibrium free-energy barrier, the effective diffusion coefficient, the mean first-passage time, and the stationary current. Particular attention is paid to the comparison between two- and three-dimensional confinement and between circular and elliptic cross sections. We show that the rod free-energy barrier depends strongly on the ratio between the rod length and the channel dimensions and that channels with the same local area can generate different barriers when their cross-sectional shapes differ. The effective longitudinal diffusivity is largest near the channel bottleneck, where confinement preferentially aligns the rod with the channel axis. Furthermore, the mean first-passage times obtained for different cross-sectional aspect ratios can largely be rationalized in terms of the rod's confinement-induced free-energy barrier. Finally, we examine how a constant longitudinal force, which may represent an electric force acting on a charged rod, modifies the passage time and stationary transport.

The remainder of the paper is organized as follows. Section~\ref{SEC:Model} introduces the channel and rod geometries, derives the confinement conditions, and develops the effective one-dimensional transport equation. Section~\ref{SEC:Evaluation} presents the free-energy barriers, effective diffusion coefficients, mean first-passage times, and force-driven currents for circular and elliptic channels. Section~\ref{sec:Conclusion} summarizes the main conclusions and discusses extensions to more general particle shapes, surface interactions, and electrostatic confinement.

\section{Modeling}\label{SEC:Model}
%%%%%%%%%%%%%%%%%%%%%%%%%%%%%%%%%%%%%%

We investigate the transport of a three-dimensional rigid rod through a three-dimensional channel which has a varying cross section of circular or elliptic shape extending the setting in~\cite{Malgaretti_2020} to three dimensions.

%%%%%%%%%%%%%%%%%%%%%%%%%%%%%%%%%%%%%%
\subsection{Geometry of the three-dimensional channel}\label{SEC:3D_channel}
\begin{figure}[h]
    \centering
    \includegraphics[width=\linewidth]{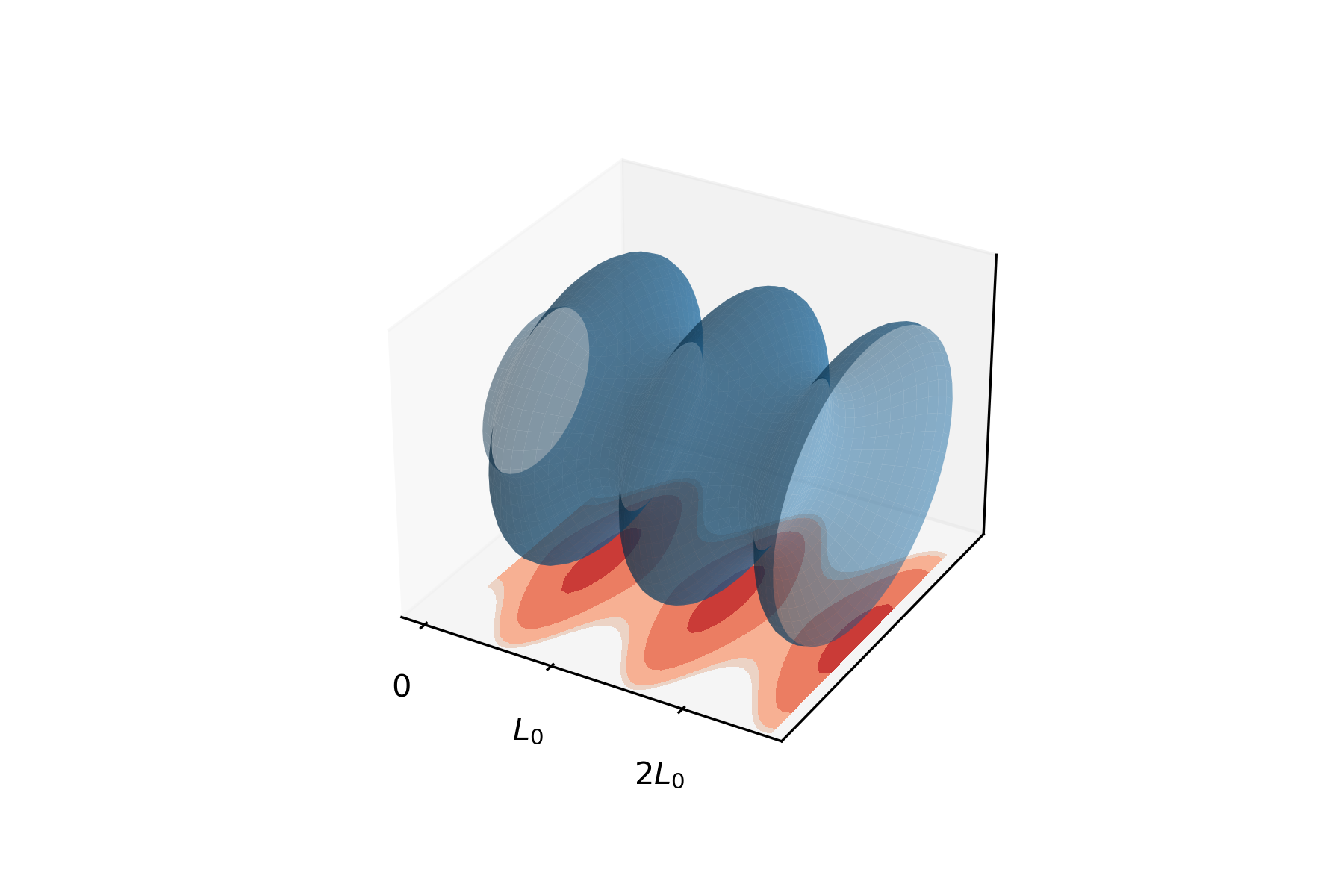}
    \caption{Three-dimensional channel with periodically varying circular cross section and projection to the $x$-$y$-plane with period $L_0$. %\st{for parameter choice $R_0=1\mu m$, $R_1=0.3\mu m$, and $L_0=2\pi \mu m$}.
    %\pa{PM: we need to remove the numbers on the axis. This should be just a cartoon. In this view we may introduce $L_0$ i.e. the lentght of the channel.}\NR{Left: old fig, Right: New fig, code is in Pics\_Ray\_new and can easily be adapted.}
    }
    \label{fig:Wavy_Channel}
\end{figure}

%%%%%%%%%%%%%%%%%%%%%%%%%%%%%%%%%%%%%%
We consider a three-dimensional periodic channel of length $L_0$ with varying circular or elliptic cross section which is extended in horizontal direction (along the $x$-axis). In the perpendicular directions ($y$- and $z$-axes), it exhibits a modulated cylindrical geometry, see illustration in Figure~\ref{fig:Wavy_Channel} for the situation of a circular cross section. 

\iffalse
For a circular cross section, we denote the altering radius of the cross section by $R(x)$. As it is advantageous to represent the corresponding circle line in cylindrical coordinates with polar angle $\varphi$ instead of using the Cartesian coordinates, we define
\begin{align}
    (x,y,z)&=(x,R(x)\cos(\varphi),R(x)\sin(\varphi)) && \text{circular cross section}.
\end{align}
We further assume that the circular cross section of the channel has an $x$-dependent radius
\begin{align}
    R(x)= R_0\left(1-R_1\cos\left(\frac{2\pi}{L_0}x\right)\right).
\end{align}
where $R_0$ denotes the average radius, i.e. width and height of the channel, and the periodic modulation with range $R_1$ and frequency~$\frac{2\pi}{L_0}$ is realized by the cosine term. 
\fi

For an elliptic cross section, we denote the varying half axes by $a(x)$ (width) and $b(x)$ (height) and define
\begin{align}
    (x,y,z)&=(x,a(x)\cos(\varphi),b(x)\sin(\varphi))\,. %&& \text{elliptic cross section}
\end{align}
The modulation of the elliptic cross section is then parametrized via the x-dependent half- axes
\begin{subequations}\label{eq:def-geom}
\begin{align}
    a(x)&= R_0\left(1-a\cos\left(\frac{2\pi}{L_0}x\right)\right)\,,\\
    b(x)&= R_0\left(1-b\cos\left(\frac{2\pi}{L_0}x\right)\right)\,,
\end{align}
\end{subequations}
with average width $R_0$ and modulation ranges $a$ and $b$, respectively. The circular cross section results by setting $R_1:=a=b$.

%%%%%%%%%%%%%%%%%%%%%%%%%%%%%%%%%%%%%%
\subsection{Geometry of the rod and confinement}\label{SEC:3D_Geometry_rod}
%%%%%%%%%%%%%%%%%%%%%%%%%%%%%%%%%%%%%%

%%%%%%%%%%%%%%%%%%%%%%%%%%%%%%%%%%%%%%
\subsubsection{Geometry of the rod}
%%%%%%%%%%%%%%%%%%%%%%%%%%%%%%%%%%%%%%

We consider a thin rod of cylindrical shape with an extension of $2L$ in its longitudinal direction and a radius of $l$ in its perpendicular direction. Moreover, we assume small aspect ratios i.e. $\tfrac{l}{L} \ll 1$. Consequently, the impact of the extension of the rod in perpendicular direction can be neglected for any further geometrical considerations. %\NR{In our numerical investigations, we nonetheless discuss \textit{isotropic} rods with $L=l$ (in appendix) and also point particles.}\\
%\pa{I do not understand? if $L=l$ the rod is a sphere isn't it?}\NR{No for $L=l$, the rod is like a fat cylinder.}

The orientation of the rod with respect to its center of mass $(x_M,y_M,z_M)$ is characterized by the Euler angles $(\theta,\Phi)$ with inclination (polar angle) $\theta\in [0,\pi]$ and azimuth $\Phi\in[0,2\pi]$. We can then represent the rod as the following line segment
\begin{align}\label{EQU:Line_rod}
\left\{    \mathbf{x} \in\IR^3 : 
\mathbf{x}  = \begin{pmatrix}
    x\\y\\z
    \end{pmatrix}
    =
    \begin{pmatrix}
    x_M\\y_M\\z_M
    \end{pmatrix}
    +t
    \begin{pmatrix}
    \sin(\theta)\cos(\Phi)\\\sin(\theta)\sin(\Phi)\\ \cos(\theta)
    \end{pmatrix} \text{ with }
-L\leq t\leq L
\right\}.
\end{align}

%\begin{remark}
We remark that here we define the inclination (polar angle)~$\theta\in [0,\pi]$ as the angle between the zenith direction ($z$-direction) and the rod orientation. This deviates from the definition of the polar angle in~\cite{Malgaretti_2020} (see illustration in Figure 1 therein) for the two-dimensional situation where the angle was measured between the horizontal axis ($x$-direction) and the rod's orientation.
%\end{remark}

%%%%%%%%%%%%%%%%%%%%%%%%%%%%%%%%%%%%%%%%%%%%%%%%%%%%%
 \subsubsection{Confinement of the rod}\label{SEC:Condition}
%%%%%%%%%%%%%%%%%%%%%%%%%%%%%%%%%%%%%%%%%%%%%%%%%%%%%

We assume that the transport of the rod is slow and that the channel width and height are varying slowly as compared to the length of the rod. As such it suffices to approximate the modulated channel by an elliptic or circular cylinder, i.e. a channel of elliptic or circular cross section with \textit{constant} width~$a(x)$ and height~$b(x)$ or \textit{constant} radius~$R(x)$, respectively. Due to the longitudinal spatial extension of the rod, it may be confined in its motion close to the boundary of this channel, i.e. if the center of mass $(x_M,y_M,z_M)$ is closer to the boundary of the cylinder than the longitudinal extension of the rod allows for:\begin{align}
    \frac{y_M^2}{(a(x)-L)^2} + \frac{z_M^2}{(b(x)-L)^2} &> 1\,. %&& \text{for the elliptic cross section}%\\
    %y_M^2+z_M^2 &>  (R(x)-L)^2 && \text{for the circular cross section}.
\end{align}
As in the two-dimensional situation~\cite{Malgaretti_2020}, we determine the condition on the angles for the case that the tip or tail intersects the upper or lower boundary of the cylinder respectively. 
In these situations, the line segment representing the rod (cf.~\eqref{EQU:Line_rod}) must intersect the ellipses/circle line:
\begin{align}
    \frac{y^2}{a^2(x)} + \frac{z^2}{b^2(x)}=1.
\end{align}
For $z_M>0$, the tip $t=L$ crossing the boundary of the channel implies
\begin{align}
    \Gamma^+ &\geq 1\,,\\
    \text{where }\Gamma^+ &\coloneqq\frac{(y_M+L\sin(\theta)\sin(\Phi))^2}{a^2(x)} + \frac{(z_M+L\cos(\theta))^2}{b^2(x)}\,, \label{EQU:confinment_3D_1}
\end{align}
and likewise for the tail $t=-L$
\begin{align}
    \Gamma^- &\geq 1\,,\\
    \text{where }\Gamma^- &\coloneqq \frac{(y_M-L\sin(\theta)\sin(\Phi))^2}{a^2(x)} + \frac{(z_M-L\cos(\theta))^2}{b^2(x)}\,, \label{EQU:confinment_3D_2}
\end{align}
In an analogous way, the constraint is computed if the rod touches the lower part of the cylinder i.e. if $z_M<0$.
For a circular cross section, i.e. $a(x)=b(x)=R(x)$, the conditions simplify to
\begin{align}
(y_M\pm L\sin(\theta)\sin(\Phi))^2 + (z_M \pm L\cos(\theta))^2& = R^2(x). \label{EQU:condition_3D}
\end{align}

%In the following we refer to these conditions as \textit{3D constraint}. \NR{Maybe make it more precise?} They are used directly in an if-else conditional for the numerical investigations as outlined in Section~\ref{SEC:Evaluation}.

%\begin{remark}
In contrast to the two-dimensional situation the conditions derived for the three-dimensional situation can in general not be resolved explicitly for minimal and maximal angles $\theta_m,\theta_M$ and $\Phi_m,\Phi_M$ specifying the confinement. 
In Appendix~\ref{SEC:Special_cases_3D} a discussion of specific geometric settings and the consistency with the two-dimensional setting as outlined in~\cite{Malgaretti_2020} is presented. 
%\end{remark}

%\subsection{Transport of an uncharged rod}

\subsection{Transport modeling via the Smoluchowski equation}
%%%%%%%%%%%%%%%%%%%%%%%%%%%%%%%%%%%%%%

%We now consider the transport of thin rods through periodically varying channels. 
The 3D Smoluchowski %(Fokker-Planck)
equation~\cite{Risken_book} for the probability distribution $\rho(t,x,y,z,\theta,\Phi)$ describing rods with mid point $x,y,z$ and orientation $\theta,\Phi$ at time $t$ reads
\begin{align}
    \partial_t\rho - \nabla\cdot \mathbf J - \nabla_{\theta,\Phi} \cdot \mathbf J_{\theta,\Phi}=0
\end{align}
with fluxes $\mathbf J$ and $\mathbf J_{\theta,\Phi}$. The fluxes account for diffusive parts as well as conservative forces and the confinement. More precisely %it holds:
\begin{align}
    \mathbf J= (J_x,J_y,J_z)=D(\nabla\rho + \rho\beta\nabla W)\,,
\end{align}
with translational diffusion matrix~$D$. As the diffusion of a rod is reduced in the direction of the minor axes of the rod (direction $\mathbf e_\theta, \mathbf e_\Phi$) according to the aspect ratio, we obtain in the frame of reference $(\mathbf e_r,\mathbf e_\theta, \mathbf e_\Phi)$ for the rod
\begin{align}
    D=D_0
    \begin{pmatrix}
    1 & 0& 0\\
    0&\tfrac{l}{L} & 0\\
    0& 0&\tfrac{l}{L}\,,
    \end{pmatrix}\,,
\end{align}
where $D_0$ denotes the diffusivity of an isotropic particle. 
Remark that due to the symmetry of the rods, the off diagonal entries are zero.

In order to determine the diffusion in the Cartesian coordinate frame, we apply a transformation of the coordinate system
\begin{align}\label{EQU:Diff_3D}
    \mathbf e_x\cdot D \mathbf e_x 
    % &
    % = \left(\sin(\theta)\cos(\phi) \mathbf e_r +  \cos(\theta)\cos(\phi) \mathbf e_\theta -\sin(\phi)\mathbf e_\phi \right) \cdot \nonumber\\
    % &\hspace{1cm} D_0
    % \begin{pmatrix}
    % 1 & 0& 0\\
    % 0&\tfrac{l}{L} & 0\\
    % 0& 0&\tfrac{l}{L}
    % \end{pmatrix}
    % \left(\sin(\theta)\cos(\phi) e_r +  \cos(\theta)\cos(\phi) e_\theta -\sin(\phi)e_\phi \right)\nonumber\\
    &= D_0\left(\sin^2(\theta)\cos^2(\Phi) + \tfrac{l}{L}\cos^2(\theta)\cos^2(\Phi) + \tfrac{l}{L}\sin^2(\Phi)\right).
\end{align}
which is consistent with the result obtained in~\cite{Malgaretti_2020} for two spatial dimensions by setting $\Phi=0$. 
%Moreover, for \textit{isotropic} rods ($l=L$), it holds
%\begin{align}
%    \mathbf e_x\cdot D \mathbf e_x 
%    = D_0\left(\sin^2(\theta)\cos^2(\phi) + \cos^2(\theta)\cos^2(\phi) + \sin^2(\phi)\right)
%   = D_0.
%\end{align}
%\end{remark}
The forcing term $\nabla W$ is scaled with the inverse thermal energy $\beta=(k_BT)^{-1}$ and accounts for the longitudinal forces $f$ as well as equilibrium conservative potentials $\phi(x,y,z,\theta,\Phi)$ and the confinement by the channel boundaries due to the spatial extension of the rods. In summary,
\begin{align}\label{equ:potential_phi}
    W (x,y,z,\theta,\Phi)=
    \begin{cases}
    \phi(x,y,z,\theta,\Phi) -fx & \text{ if } \,\Gamma^+\!\!<1 \text{\,\,and\,\,} \Gamma^-\!\!<1%, cf.~\eqref{EQU:confinment_3D_1},~\eqref{EQU:confinment_3D_2}
    \\
    \infty & \text{ else }\,.
    \end{cases}
\end{align}
Finally, the rotational components are given by 
\begin{align}
    \mathbf J_{\theta,\Phi}= (J_\theta,J_\Phi)= D_{\theta,\Phi}\nabla_{\theta,\Phi}\rho\,,
\end{align}
with the rotational diffusion matrix %$D_{\theta,\Phi}$ 
\begin{align}
    D_{\theta,\Phi} =
    \begin{pmatrix}
    D_\theta & 0\\
    0& D_\Phi
    \end{pmatrix}.
\end{align}

%%%%%%%%%%%%%%%%%%%%%%%%%%%%%%%%%%%%%%%%%%%%%%%%%%%
 \subsubsection{Fick-Jacobs approximation}\label{Fick_Jacobs_App}
%%%%%%%%%%%%%%%%%%%%%%%%%%%%%%%%%%%%%%%%%%%%%%%%%%%

In order to simplify the model, we assume that the translation along the channel axis ($x$) is slow, such that an equilibrium is obtained along $(y,z,\theta,\Phi)$, i.e.
\begin{align}
    J_y = J_z = J_\theta = J_\Phi = 0\,.
\end{align}
Consequently
\begin{align}\label{EQU:Smoluchowski_3D_start}
    \partial_t\rho - \partial_x\left((\mathbf e_x\cdot D\mathbf e_x)(\partial_x\rho + \rho\beta\partial_x W)\right) = 0\,,
\end{align}
with $\mathbf e_x\cdot D\mathbf e_x$ from~\eqref{EQU:Diff_3D} and $W$ depending on $(x,y,z,\theta,\Phi)$ as discussed above.
We apply the Fick-Jacobs approximation~\cite{zwanzig,Burada_Review}
\begin{align}\label{EQU:Rep_rho}
\rho(t,x,y,z,\theta,\Phi)=p(t,x)\frac{e^{-\beta W(x,y,z,\theta,\Phi)}}{e^{-\beta A(x)}}\,,
\end{align}
with the local equilibrium free energy
\begin{align}
\beta A(x) = -\ln \left(\frac{1}{\pi R_0^2} \int_0^\pi\int_0^{2\pi}\int_{-\infty}^\infty\int_{-\infty}^\infty   e^{-\beta W(x,y,z,\theta,\Phi)} dydzd\theta d\Phi \right)\,.
\end{align}
%and normalization constant~$K$ \NR{$=\pi R_0^2$ see \eqref{eq:def-A}?}.
We rewrite the individual terms of the transport equation~\eqref{EQU:Smoluchowski_3D_start} as follows: First, for the time derivative in~\eqref{EQU:Smoluchowski_3D_start}, we have
 \begin{align}%\label{EQU:Rep_rho_2D_time_derivative}
     \partial_t \rho = (\partial_t p) \frac{e^{-\beta W}}{e^{-\beta A}}.
 \end{align}
Second, the horizontal flux term in~\eqref{EQU:Smoluchowski_3D_start} is investigated. Due to~\eqref{EQU:Rep_rho} and the product and quotient rule for differentiation, we obtain
\begin{align}
    \partial_x\rho + \rho\beta\partial_x W 
    &= \partial_x p \frac{e^{-\beta W}}{e^{-\beta A}} 
    +  p \partial_x\left(\frac{e^{-\beta W}}{e^{-\beta A}} \right)
    + p \frac{e^{-\beta W}}{e^{-\beta A}}\beta \partial_x W\,,\\
    % &= \partial_x p \frac{e^{-\beta W}}{e^{-\beta A}} 
    % + p \frac{e^{-\beta A}\partial_x e^{-\beta W} - e^{-\beta W}\partial_x e^{-\beta A} }{\left(e^{-\beta A}\right)^2}
    % + p\frac{e^{-\beta W}}{e^{-\beta A}}\beta \partial_x W\\
    % &= \partial_x p \frac{e^{-\beta W}}{e^{-\beta A}} 
    % + p\left( -\beta\frac{ e^{-\beta W}\partial_x W}{e^{-\beta A}}
    % +\beta \frac{\beta e^{-\beta W} e^{-\beta A}\partial_x A }{\left(e^{-\beta A}\right)^2}
    % + \frac{e^{-\beta W}}{e^{-\beta A}}\beta \partial_x W\right)\\
    &= \left(\partial_x p + p\beta \partial_x A\right) \frac{e^{-\beta W}}{e^{-\beta A}}.
\end{align}
In summary, it holds
\begin{align}
(\partial_t p) \frac{e^{-\beta W}}{e^{-\beta A}}
-\partial_x\left((\mathbf e_x\cdot D\mathbf e_x) \left(\partial_x p + p\beta \partial_x A\right) \frac{e^{-\beta W}}{e^{-\beta A}}\right) =0. 
\end{align}
We then calculate the average with respect to the variables $(y,z,\theta,\Phi)$ and since the main unknown $p$ only depends on $(t,x)$, we obtain for the evolution term
 \begin{align}
     \int_0^\pi\int_0^{2\pi}\int_{-\infty}^\infty\int_{-\infty}^\infty \partial_t \rho dydzd\theta d\Phi &= \partial_t p \int_0^\pi\int_0^{2\pi}\int_{-\infty}^\infty\int_{-\infty}^\infty\frac{e^{-\beta W}}{e^{-\beta A}} dydzd\theta d\Phi=\partial_t p
     %= K\partial_x\left(\mathcal{D}\left(\partial_x p + p\beta \partial_x A\right)\right)
 \end{align}
 and for the flux term
 \begin{align}
    \partial_x &\left[\int_0^\pi\int_0^{2\pi}\int_{-\infty}^\infty\int_{-\infty}^\infty J_x dydzd\theta d\Phi\right] =\nonumber\\
    &=\partial_x\left[\left(\partial_x p + p\beta \partial_x A\right)\int_0^\pi\int_0^{2\pi}\int_{-\infty}^\infty\int_{-\infty}^\infty (\mathbf e_x \cdot D\mathbf e_x) \frac{e^{-\beta W}}{e^{-\beta A}} dzdyd\Phi d\theta \right]\nonumber\\
&=   \partial_x\left(\mathcal{D}(x)\left(\partial_x p + p\beta \partial_x A\right)\right)
\end{align}
with an effective diffusion
\begin{align}\label{EQU:3D_Diff}
    \mathcal{D}(x) = %\tfrac{1}{K} 
    \int_0^\pi\int_0^{2\pi}\int_{-\infty}^\infty\int_{-\infty}^\infty \mathbf e_x \cdot D \mathbf e_x \frac{e^{-\beta W}}{e^{-\beta A}} dydzd\theta d\Phi\,.
\end{align}
With~\eqref{EQU:Diff_3D}, we write $\mathcal{D}(x)=\frac{\mathcal{D}_{nom}}{\mathcal{D}_{denom}}$, where
\begin{align}
\mathcal{D}_{nom} &\coloneqq %\tfrac{1}{K} 
\int_0^\pi\int_0^{2\pi}\int_{-\infty}^\infty\int_{-\infty}^\infty \mathbf e_x\mathbf  \cdot D \mathbf e_x e^{-\beta W} dydzd\theta d\Phi ,
\label{EQU:Diff_nom}\\
\mathcal{D}_{denom} &\coloneqq e^{-\beta A}  .
\label{EQU:Diff_denom}
\end{align}
We recall that while the free energy is likely to be a function of $L/R_0$ the diffusion coefficient is not. This is due to the term $l/L$ in Eq.~\eqref{EQU:Diff_3D}. 
In summary, we obtain the same averaged transport equation as for the two-dimensional situation (cf.~\cite{Malgaretti_2020}), but with a different definition of the effective diffusion coefficient and free energy:
 \begin{align}
    \partial_t p(x,t) = \partial_x\left(\mathcal{D}(x)\left(\partial_x p(x,t) + p(x,t)\beta \partial_x A(x)\right)\right).
    \label{eq:FJ-rod_3D}
 \end{align}
 
 %%%%%%%%%%%%%%%%%%%%%%%%%%%%%%%%%%%%%%%%%%%%%%%%%%%
 \subsubsection{Steady state  solution}
 %%%%%%%%%%%%%%%%%%%%%%%%%%%%%%%%%%%%%%%%%%%%%%%%%%%

The steady state solution of Eq.~\eqref{eq:FJ-rod_3D} reads
 \begin{align}
  \partial_x\left(\mathcal{D}(x)\left(\partial_x p(x) + p(x)\beta \partial_x A(x)\right)\right)=0\,,
 \end{align}
and it can be solved by integration and the solution method \textit{separation of variables} as follows: After integration and introducing the integration constant~$J$, we obtain the following inhomogeneous first-order differential equation 
 \begin{align}\label{EQU:ODE_p_2D}
  \partial_x p(x) + p\beta \partial_x A (x) =\frac{J}{\mathcal{D}(x)}.
 \end{align}
 Separation of variables first yields the homogeneous solution
 \begin{align}
  p^{hom}(x)= %\Pi e^{-\int\beta \partial_x A(x)} = 
  \Pi e^{-\beta A(x)}, \quad \Pi\in\mathbb{R}\,.
 \end{align}
%to $\partial_x p^{hom}(x)=-\beta \partial_x A(x) p^{hom}(x)$ since $\beta$ is a constant.
Then, the method \textit{variation of the constant} with the ansatz
\begin{align}
  p^{inhom}(x)= \Pi(x) e^{-\beta A(x)}\,,
 \end{align}
yields 
% the following differential equation for the prefactor~$\Pi(x)$ (by means of~\eqref{EQU:ODE_p_2D})
% \begin{align}
%     \partial_x \Pi(x) e^{-\beta A(x)} = \frac{J}{\mathcal{D}(x)}
% \end{align}
% i.e.
\begin{align}
    \Pi(x) = \int_0^x \frac{J}{\mathcal{D}(x')}e^{\beta A(x')} dx'.
\end{align}
The general solution as superposition of the homogeneous solution $p^{hom}$ and particular solution $p^{inhom}$ then reads
   \begin{align}
  p(x)= p^{hom}(x) + p^{inhom}(x)= e^{-\beta A(x)} \left( \Pi + \int_0^x \frac{J}{\mathcal{D}(x')} e^{\beta A(x')} dx'\right)\,,
 \end{align}
 with
 \begin{align}\label{EQU_flux3D}
     \Pi&= \frac{\int_0^{L_0}\frac{J}{\mathcal{D}(x')} e^{\beta A(x')} dx' e^{-\beta A(L)}}{e^{-\beta A(0)} - e^{\beta A(L_0)}}
     = J\frac{\int_0^{L_0}\frac{1}{\mathcal{D}(x')} e^{\beta A(x')} dx' }{e^{-\beta (A(0)- A(L_0))} - 1} =: J\Pi_0,\\
     J &= \left( \int_0^{L_0}e^{-\beta A(x)} \left(\Pi_0 + \int_0^x \frac{1}{\mathcal{D}(x')} e^{\beta A(x')} dx'\right) dx\right)^{-1}.
 \end{align}
For weak driving forces, we can extract the effective channel permeability for the rod within the linear response regime as~\cite{Malgaretti_2023}
\begin{align}
    \mu = \dfrac{2\beta L_0}{\int_0^{L_0}e^{-\beta A_{eq}(x)} dx \int_0^{L_0} \frac{1}{\mathcal{D}_{eq}(x)} e^{\beta A_{eq}(x)}  dx}\,,
\end{align}
where $A_{eq}$ and $\mathcal{D}_{eq}$ indicate that they have been computed at equilibrium, i.e. with $f=0$. In the case of a constant section channel and for a spherical colloid, with radius $l$ and diffusion coefficient $D_0$, the permeability simplifies to
\begin{align}
    \mu_0 = 2\beta D_0/L_0 \,.
\end{align}
\iffalse
Then with the choices $\beta=1, f=1$, we obtain
\begin{align}\label{EQU:Time}
    T(x)=\frac{1}{D_0}\int_0^x \frac{e^{-x'}}{\mathcal{D}_{denom}} \left(\int_0^{x'} \frac{e^{z}\mathcal{D}_{denom}^2(z)}{\mathcal{D}_{nom}(z)} dz\right) dx'
\end{align}
with denominator~$\mathcal{D}_{denom}$ and nominator~$\mathcal{D}_{nom}$ of the diffusion coefficient~$\mathcal{D}$. 

A a remark note that~\eqref{EQU:Time} holds for the two- as well as three-dimensional situation. However, the definition of $\mathcal{D}_{denom}$ and~$\mathcal{D}_{nom}$ differ.
\fi

%%%%%%%%%%%%%%%%%%%%%%%%%%%%%%%%%%%%%%%%%%%%%%%%%%%%%
\section{Results}\label{SEC:Evaluation}
%%%%%%%%%%%%%%%%%%%%%%%%%%%%%%%%%%%%%%%%%%%%%%%%%%%%%
%%The numerical evaluations turned out to be very sensitive and were carried out carefully with python. Due to the discontinuity of the integrand (contributing to the integral only if the constraint does not apply), it turned out that the integration toolbox provided in python, namely the routine \textit{dblquad} and \textit{nquad} were not capable of the integration and neither was a Monte Carlo approach. As a consequence, we applied a simple decision rule (whether the integrand is 0 or 1) combined with a lower Riemann approximation to the remaining integral. That means that the integral is approximated by a sum of cuboids with constant base area (related to the chosen incremental step sizes) and varying height. 
In this section, the free energy barrier, effective diffusion, mean first passage time, and flux are studied as dependent on aspect ratio and channel geometry. All calculations were carefully carried out in python applying a simple decision rule (whether the integrand is 0 or 1) combined with a lower Riemann approximation to the remaining integral.

\subsection{Equilibrium free energy barrier}

We consider the potential 
 \begin{align}
    W (x,y,z,\theta,\Phi)=
    \begin{cases}
    0 &  \begin{cases}\Gamma^+<1\\ \Gamma^-<1\end{cases} \\
    \infty & \text{else}\,,
    \end{cases}
\end{align}
i.e. focusing on the confinement while neglecting longitudinal and further forces.
%cf.~\eqref{equ:potential_phi} with $\phi=0$ and \NR{$f=0$}.
Then the equilibrium free energy barrier is given as
\begin{align}
A(x)=- k_BT\ln\left(\tfrac{1}{\pi R_0^2}\int_0^\pi\int_0^{2\pi}\int_{-\infty}^\infty\int_{-\infty}^\infty e^{-\beta W(x,y,z,\theta,\Phi)}dydzd\theta d\Phi\right).
\label{eq:def-A}
\end{align}
%and $\Delta A_{eq,3D} = A(R_{min},f=0)-A(R_{max},f=0)$. 
For an ideal gas composed of point-particles, %\NR{we call it point particle below}, 
Eq.~\eqref{eq:def-A} reduces to 
\begin{align}
A_{gas}(x)=-k_BT\ln \frac{R^2(x)}{R_0^2} =-2 k_BT\ln \frac{R(x)}{R_0} 
\end{align}
in $3D$ which differs from the $2D$ setting by a factor of 2, cf. ~\cite{Malgaretti_2020}. 

\begin{figure}[ht]
    \centering
    \includegraphics[scale = 0.32]{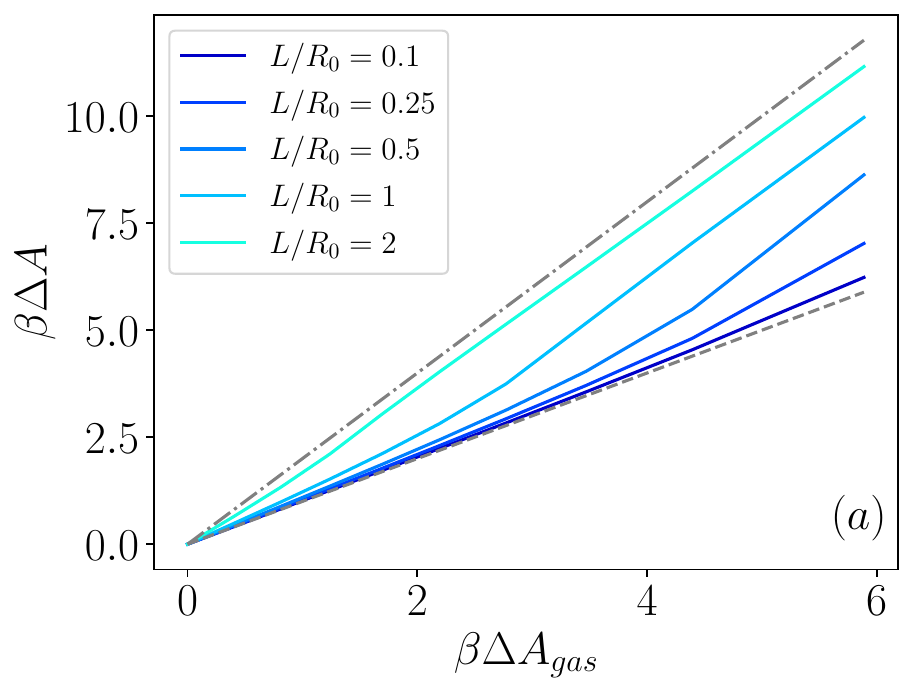}
    \includegraphics[scale = 0.32]{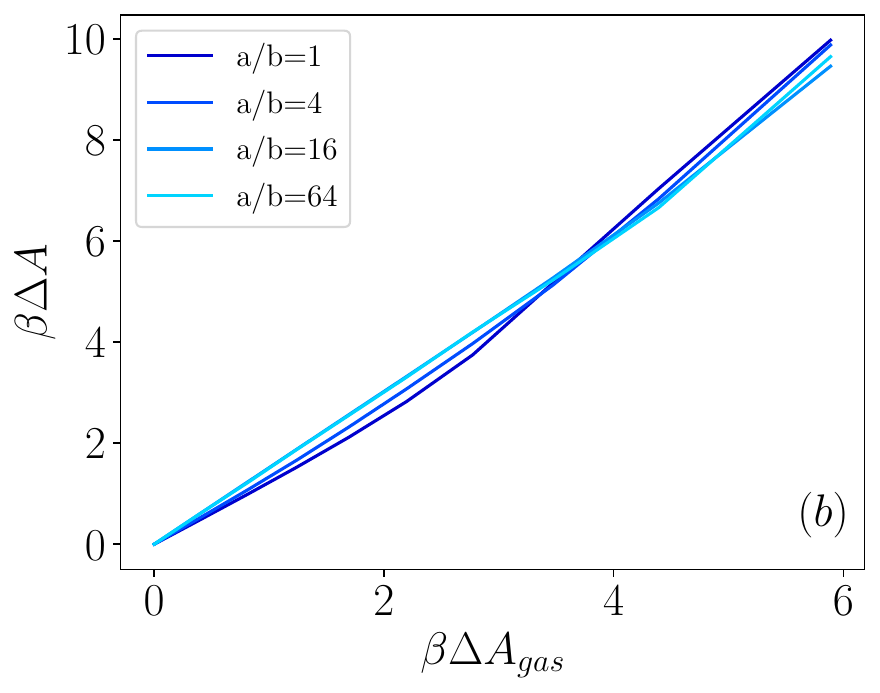}
    \includegraphics[scale = 0.32]{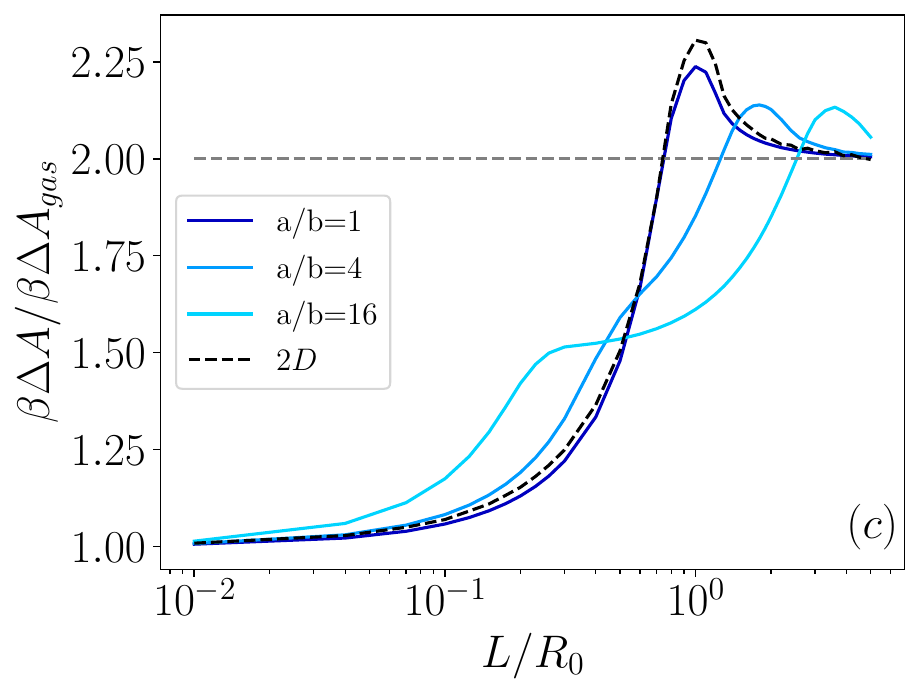}
    \caption{(a): Equilibrium free energy barrier $\beta\Delta A$ over gas free energy $\beta\Delta A_{gas}$ for $L/R_0$ as reported in the legend and circular channel cross-section, $a/b=1$. Dotted line: $\beta\Delta A_{gas}$; dash-dotted line: $2\beta\Delta A_{gas}$ (b): Equilibrium free energy barrier $\beta\Delta A$ over gas free energy $\beta\Delta A_{gas}$ for $L/R_0=0.3$ and aspect ratio of the elliptic channel cross section as reported in the legend.
    (c): Equilibrium free energy barrier $\beta\Delta A$ normalized by the gas free energy $\beta\Delta A_{gas}=0.5$, as a function of $L/R_0$ where $R_0$ is the average channel section defined in Eq.~\eqref{eq:def-geom}, and aspect ratio of the elliptic channel cross section as reported in the legend. The black dashed line is for the 2D case, and the grey horizontal line is a guide for the eye.}
    \label{fig:DA}
\end{figure}

The equilibrium free energy barrier in dependence of the ideal gas equilibrium free energy barrier for varying rod lengths and channel cross sections are depicted in Figure~\ref{fig:DA}.
As expected, Fig.~\ref{fig:DA}a shows that the free energy barrier $\Delta A$ of rods much shorter than the average channel radius, $L/R_0\ll 1$, approaches that of a point particle (dotted line), at least for $\beta \Delta A_{gas}\lesssim 1$ and increasing with decreasing rod length,  whereas for $L \gtrsim R_0$ Fig.~\ref{fig:DA}a shows $\Delta A \simeq 2 \Delta A_{gas}$ (dash-dotted line). Next we investigate the dependence of the free energy barrier upon varying the aspect ratio of the elliptic cross-section. As shown in Fig.\ref{fig:DA}b
 $\Delta A$ is larger than $\Delta A_{gas}$ for larger aspect ratios at smaller corrugations (i.e. $\beta\Delta A_{gas}\leq 4$), whereas the opposite holds at larger values of $\Delta A_{gas}$. %\NR{Still I am wondering why?}. 
 Finally, Fig.~\ref{fig:DA}c shows that the dependence of $\Delta A$ on $L/R_0$ %\NR{Here and in the lines below $r_0=R_0$?}
 is quite sensitive to the aspect ratio. For circular cross-sections, $a/b =1$, the 3D data resembles the 2D one (dashed line in Fig.\ref{fig:DA}c). In contrast, when $a/ b\neq 1$ the behavior changes significantly, showing a flattening in the curve at $L/R_0\simeq b_{min}$, i.e. when the rods matches the minimum of the smallest axis of the elliptic cross-section. Similarly, the maximum is shifted at $L/R_0\simeq a_{min}$ namely, when $L$ matches the maximum axis at the overall bottleneck. 

%\NR{What can we say about 3b? Can we explain the switch of curves?}
%\pa{FOr the moment I have simply described what we see I do not have yet an idea why it is so...}

\subsection{Effective diffusion}
\begin{figure}[h]
    \centering
    \includegraphics[width=0.7\linewidth]{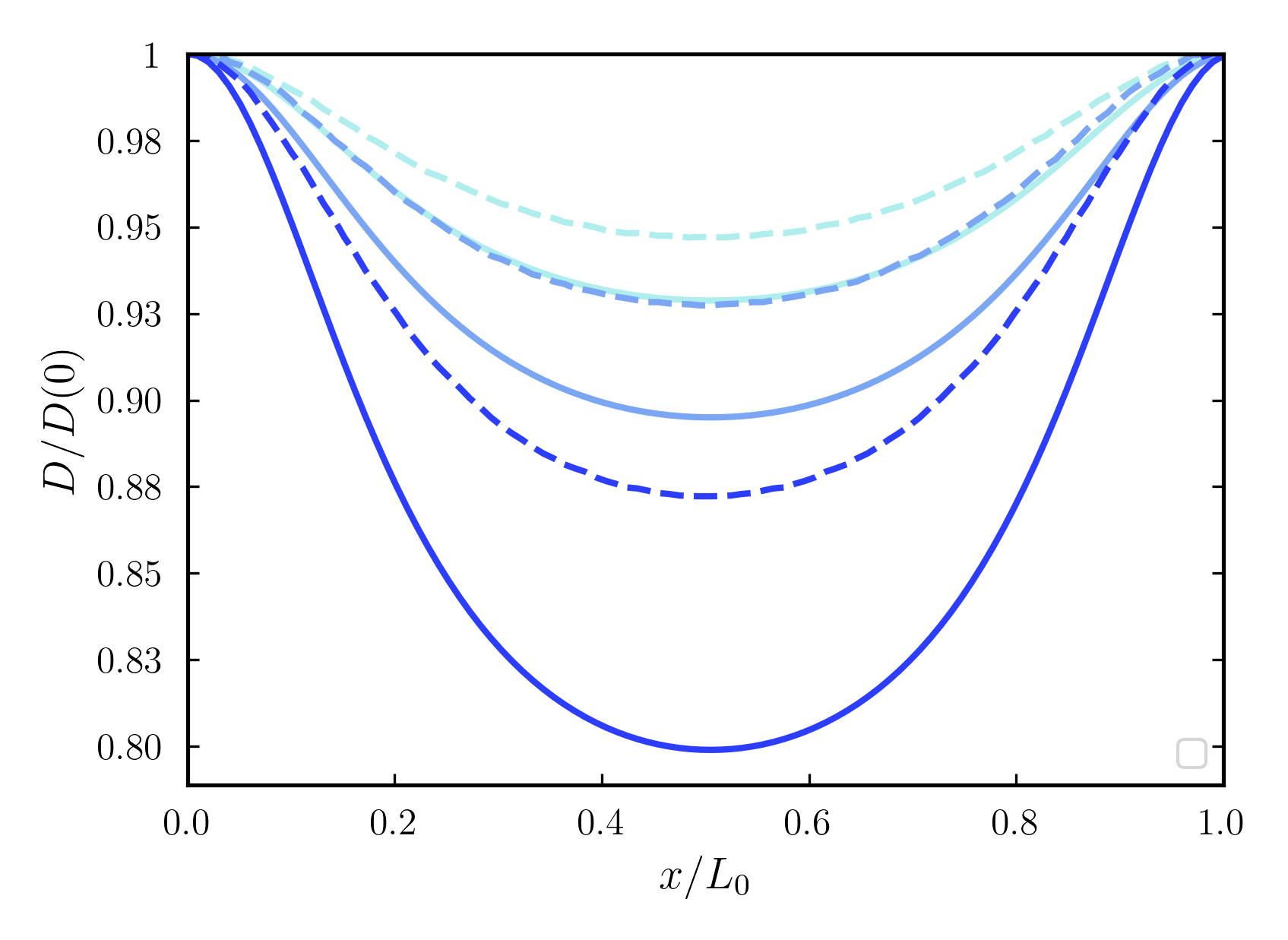}
    \caption {%\NR{changed caption such that L in dep. of R0}
    Simulated effective diffusion $D$ in two-dimensional (dashed lines) and three-dimensional setting (solid lines) of a thin rod with varying rod lengths $L=R_0/3$ (cyan), $L=R_0/2.5$ (light blue), and $L=R_0/1.875$ (blue) over normalized position $x/L_0$. Normalization with respect to the maximal value $D(0)$.}
    \label{fig:D_3D_Full}
\end{figure}

We compare the diffusion coefficient for the two-dimensional situation as derived in Ref.~\cite{Malgaretti_2020} and the general three-dimensional situation as derived in Sec.~\ref{Fick_Jacobs_App}. As shown in  Fig.~\ref{fig:D_3D_Full}, for both the two-dimensional and the three-dimensional situation, the diffusion is maximal at the bottlenecks of the channel, where the rod is confined the most and hence it is mostly aligned with the channel axis. Moreover, the diffusion strongly depends on the rod length and larger rod lengths result in a larger spread of the diffusion. In comparison to the two-dimensional situation, this effect is even more pronounced in the three-dimensional setting.

%%%%%%%%%%%%%%%%%%%%%%%%%%%%%%%%%%%%%%%%%%%%%%%%%%%%%

\subsection{Mean First Passage Time (MFPT)}

%\NR{Should we mention the agreement with analytical solutions for $f=0$ and $f=1$ and comparison 2D/3D in the appendix?}

The dimensional reduction of the problem to Eq.~\eqref{eq:FJ-rod_3D} allows us to identify the local effective driving force and the local effective diffusion coefficient. Accordingly, we can use the information to predict the Mean First Passage Time (MFPT) of a rod across a corrugated channel. In fact, the MFPT, $t(x)$, over a channel period is determined by 
\begin{align}
    -\left(\beta \partial_x A(x)\right)\partial_x t(x) + \partial_x^2 t(x) = -\frac{1}{\mathcal{D}(x)}\,,
\end{align}
with the boundary conditions on the channel entry $x=0$ and final position $x_f$
\begin{align}
    \partial_x t(x)|_{x=0} = 0\,,\label{eq:BC1}\\
    t(x=x_f) = 0\,.\label{eq:BC2}
\end{align}
Defining $T(x) = \partial_x t(x)$, after integration and using Eq.~\eqref{eq:BC1} the solution reads
\begin{align}
    T(x)=-e^{\beta A(x)} \left(\int_0^{x} \frac{e^{-\beta A(z)}}{\mathcal{D}(z)} dz\right) \,,
\end{align}
from which, integrating and using Eq.~\eqref{eq:BC2}, the MFPT can be extracted as
\begin{align}\label{eq:MFPT}
    t(x,x_f) = \int_0^{x_f} e^{\beta A(x')} \left(\int_0^{x'} \frac{e^{-\beta A(z)}}{\mathcal{D}(z)} dz\right) dx' - \int_0^x e^{\beta A(x')} \left(\int_0^{x'} \frac{e^{-\beta A(z)}}{\mathcal{D}(z)} dz\right) dx'\,.
\end{align}
In particular we have that the MFPT from $x=0$ to $x=x_f$ reads
\begin{align}
    t_0(x_f) = t(x=0,x_f)=\int_0^{x_f} e^{\beta A(x')} \left(\int_0^{x'} \frac{e^{-\beta A(z)}}{\mathcal{D}(z)} dz\right) dx'\,,
\end{align}
and the MFPT from $x=0$ to $x_f=L_0$ reads
\begin{align}\label{eq:MFPT-L0}
    t_0(L_0)  =\int_0^{L_0} e^{\beta A(x')} \left(\int_0^{x'} \frac{e^{-\beta A(z)}}{\mathcal{D}(z)} dz\right) dx'\,,
\end{align}
\begin{figure}
%\centering
\includegraphics[width=0.33\linewidth]{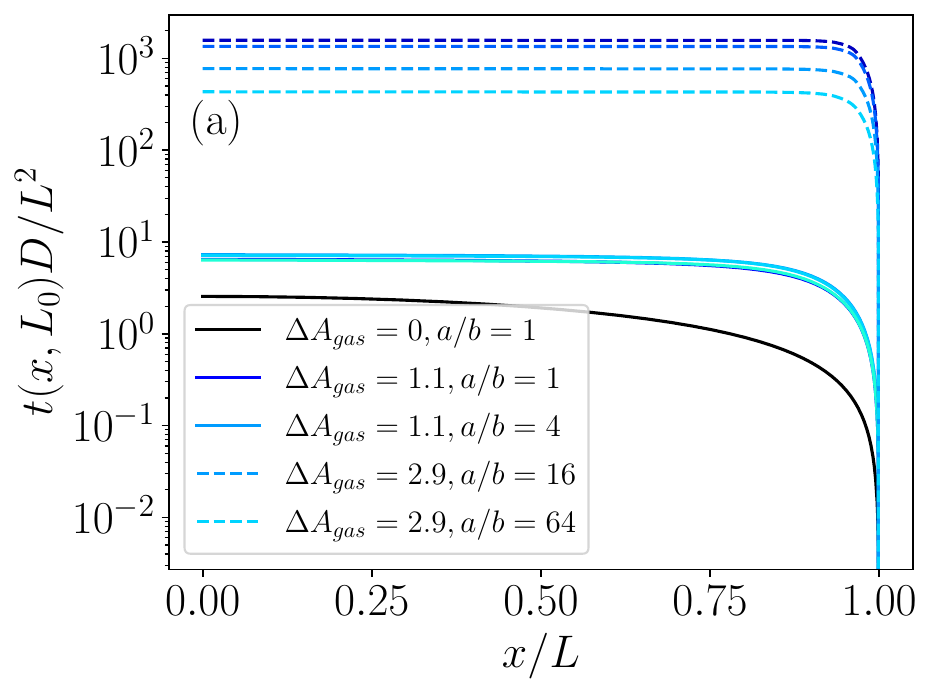}
\includegraphics[width=0.33\linewidth]{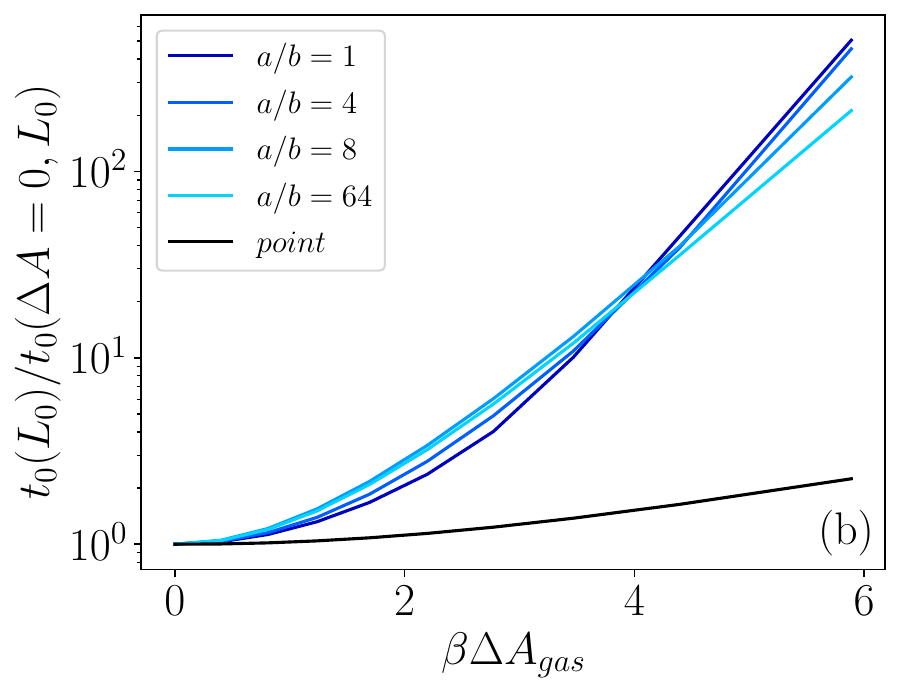}
\includegraphics[width=0.33\linewidth]{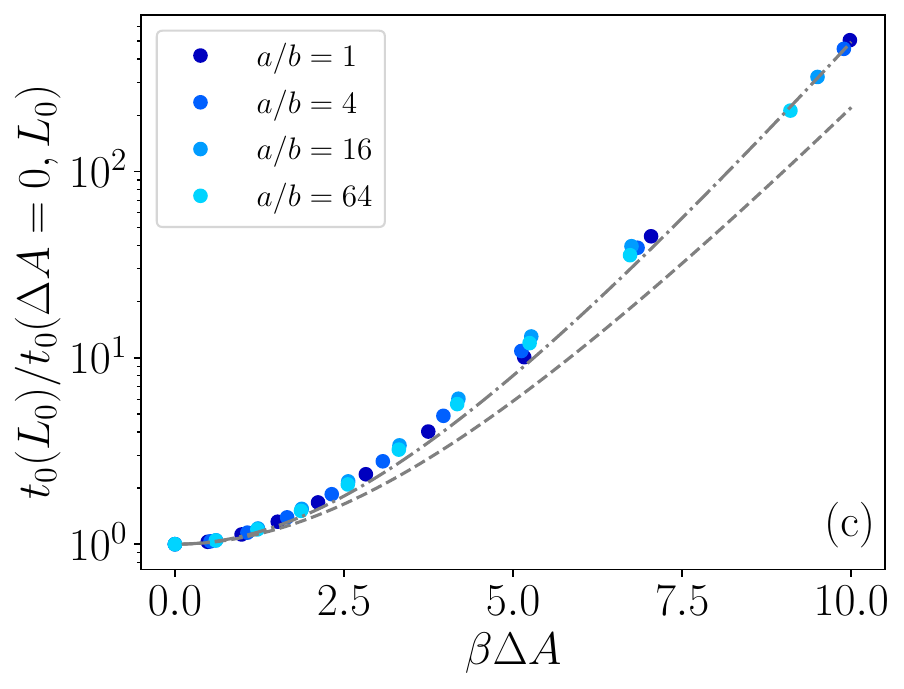}
\caption{%\NR{Can you maybe enlarge labels and axes' names?} 
(a): $t(x,x_f)$, see Eq.~\eqref{eq:MFPT}, of a rod across a corrugated channel with $L/R_0=0.3$, $L_0/R_0 = 12$, $l/L=0.1$, and $\Delta A_{gas}$ as reported in the figure. (b): $t_0(L_0)$, see Eq.~\eqref{eq:MFPT-L0}, of a rod across a corrugated channel with $L/R_0=0.3$, $L_0/R_0 = 12$, $l/L=0.1$, as a function of  $\Delta A_{gas}$ as reported in the figure for the different aspect ratios (see the legend). The black line is the $t_0$ for a point particle in the same channel. (c): same data as in panel (b) as a function of the rod free energy barrier $\Delta A$.
\label{fig:T1}
}
\end{figure}
Fig.~\ref{fig:T1}(a) shows the dependence of $t$ on the initial position, $x$. Interestingly, while for smaller values of $\Delta A_{gas}$ the MFPT is not really sensitive to the aspect ratio, it is for larger magnitudes of $\Delta A_{gas}$. In the following, we focus on the MFPT for the rod to cross the full channel, i.e. we focus on $t_0$ whereas the dependence of the MFPT on the end point is discussed in the Appendix~\ref{sec:MFPT-xf}. 
As shown in Fig.\ref{fig:T1}(b)
, the MFPT across the full channel grows upon increasing the magnitude of $\Delta A_{gas}$ and rods shows a significant enhancement of the MFPT as compared to point particles. However, the dependence on the aspect ratio is quite puzzling. In fact,  higher aspect ratios lead to longer passage times for weaker corrugations, whereas at larger corrugations they lead to a faster crossing as compared to the circular case.  
In order to rationalize this behavior, we plot $t_0$ as a function of the free energy barrier of the rod defined as the difference in the free energy at the channel waste and at the channel bottleneck. 
Interestingly, Fig.\ref{fig:T1}(c) shows that once represented as  function of $\Delta A$ the data shown in Fig.\ref{fig:T1}(b) collapse onto a master curve. In particular, the dashed line in \ref{fig:T1}(c) has been obtained by approximating the free energy profile, $A(x)$ as piece-wise linear which then gives the following closed formula for $t_0$~\cite{Malgaretti2022} 
\begin{align}
    t_0(L_0) = 2\frac{L_0^2}{D}\frac{\cosh(\beta\Delta A)-1}{(\beta\Delta A)^2}\,,
    \label{eq:MFPT-appr}
\end{align}
whereas for the dot-dashed line the free energy barrier has been fitted manually by $\Delta A \rightarrow 1.1 \Delta A$. %\NR{did you do a mean square fit or similar as most of the points are still above the line? }\\
%\pa{I just fit it by hand...}\\
The collapse of the data in Fig.\ref{fig:T1}(c) is quite remarkable and it confirms the crucial role played by the free energy profile, $A(x)$, and in particular by the free energy barrier, $\Delta A$ in the dynamics of confined rods. Indeed, since Eq.~\eqref{eq:MFPT-appr} does not account for anisotropies in the diffusion coefficient, the collapse of the data shown in Fig.\ref{fig:T1}(c) shows that the anisotropy in the diffusion coefficient is not playing a significant role in the overall MFPT. Finally, the collapse in Fig.\ref{fig:T1}(c) provides an efficient way to compute the MFPT. In fact, while  the free energy barrier, $\Delta A$, can be computed quite quickly (it only requires the free energy at two distinct points) computing~$t$ requires the full free energy profile, $A(x)$, which is more computationally demanding. 
\subsection{External forces}
\begin{figure}
    \centering
    \includegraphics[width=0.45\linewidth]{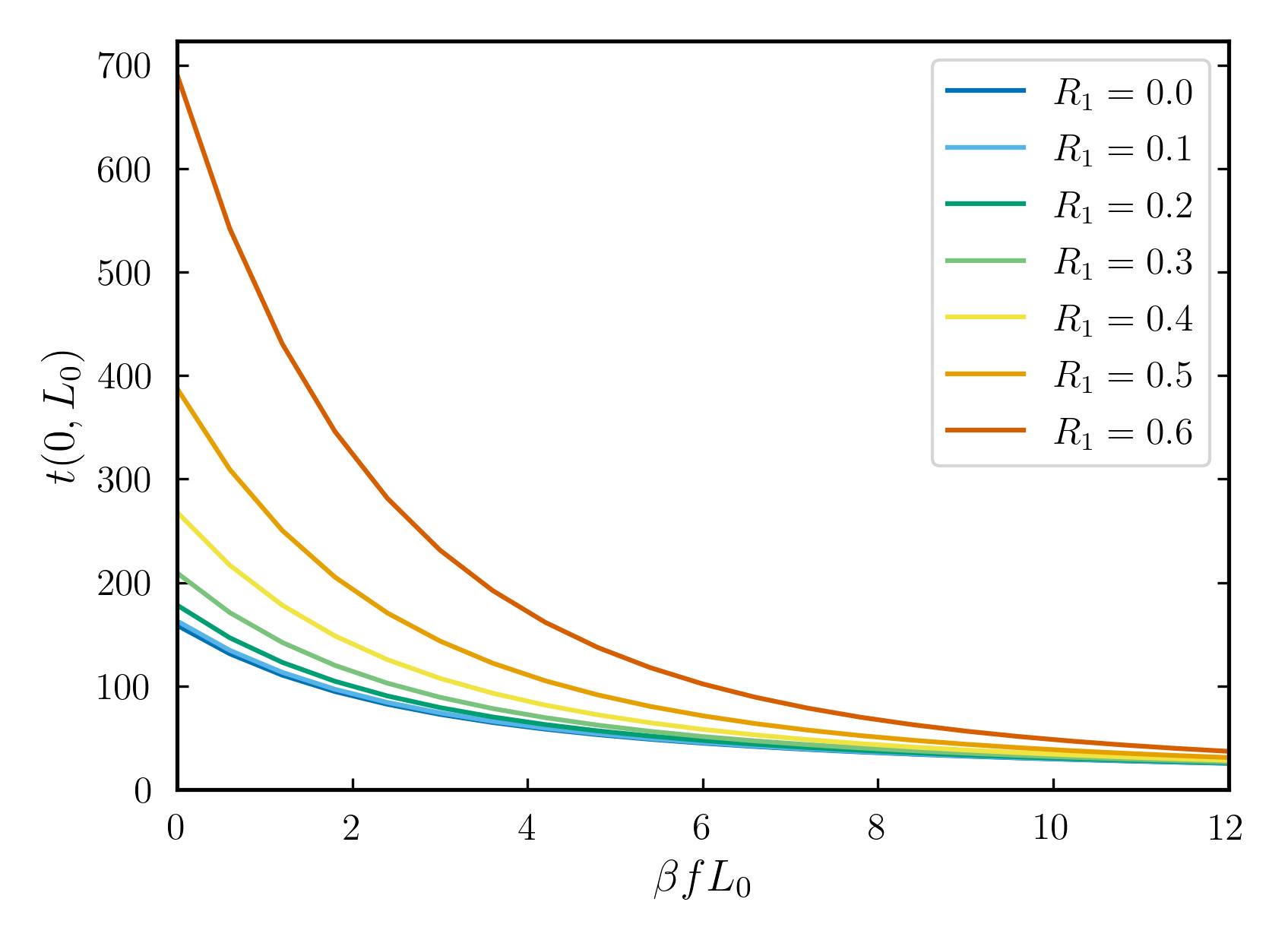}
    \includegraphics[width=0.45\linewidth]{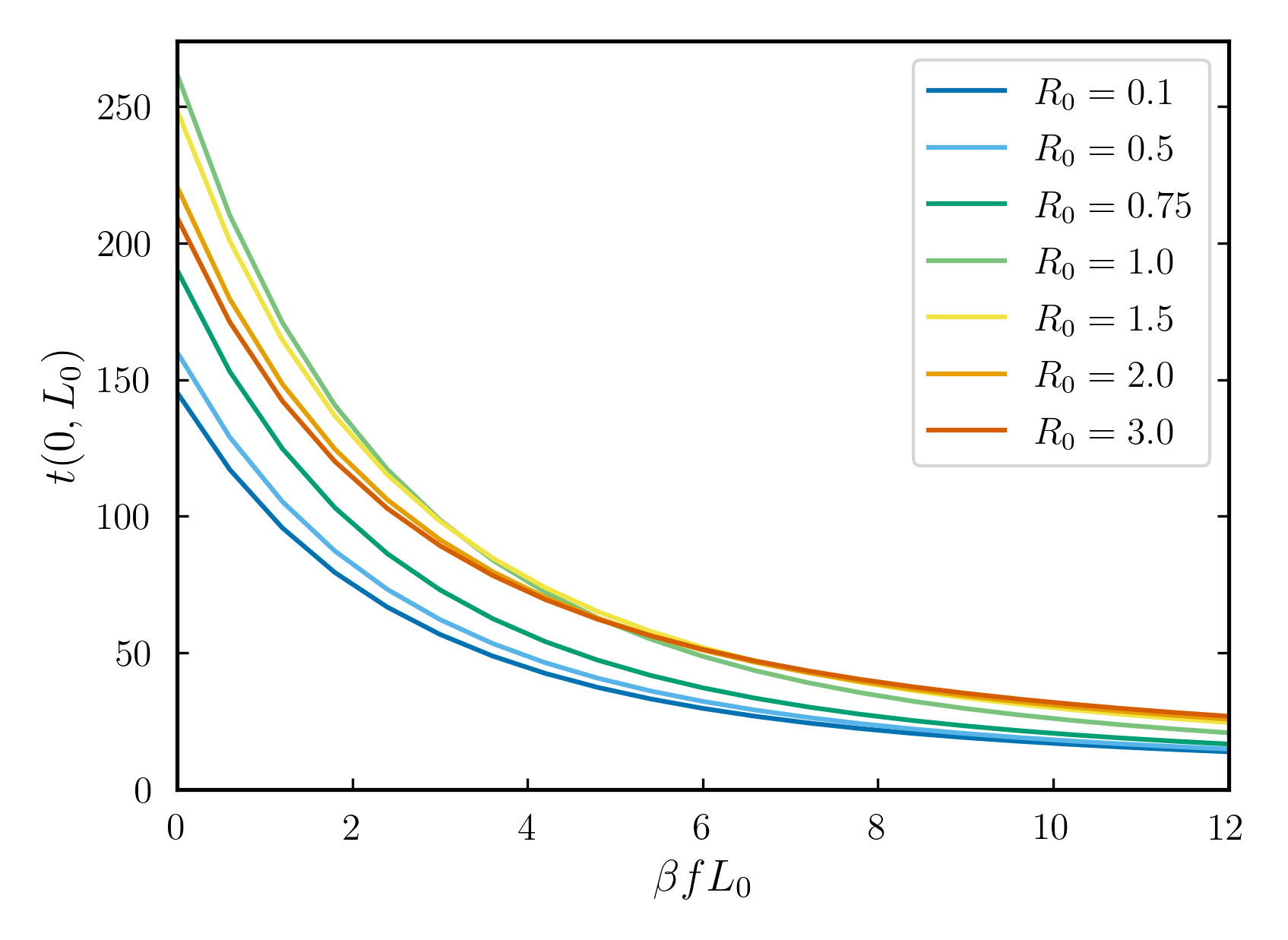}
    \caption{
    Left: End time as dependent on external force for fixed $R_0$ and varying $R_1$. Right: End time as dependent on external force for fixed $R_1$ and varying $R_0$.} %\NR{Maybe delete the following as it is described in the main text: (nonlinear as explained by $A$, cf. Fig. c)}}
    \label{fig:MFPT-force}
\end{figure}

Finally, we investigate the case in which an external force is applied to the center of mass of the rod. At first, we focus on the MFPT. As shown in Fig.\ref{fig:MFPT-force}(a) the MFPT decreases upon increasing the force and, also in the presence of the external force, it increases upon enlarging the corrugation of the channel (see also the dependence of the MFPT on the final position in Appendix~\ref{sec:MFPT-xf}). An intriguing effect appears when fixing the corrugation of the channel and varying the average channel section. As shown in Fig.\ref{fig:MFPT-force}(b), in this case there is a non-monotonic dependence of the MFPT on the corrugation of the channel (longest MFPT for $R_0=1$ under small forces.). This is due to the non-trivial dependence of the free energy barrier, $\Delta A$, on the ratio $L/R_0$ between the length of the rod and the average channel section shown in Fig.\ref{fig:DA}(c).
\begin{figure}[h]
    \centering
    \includegraphics[width=0.5\linewidth]{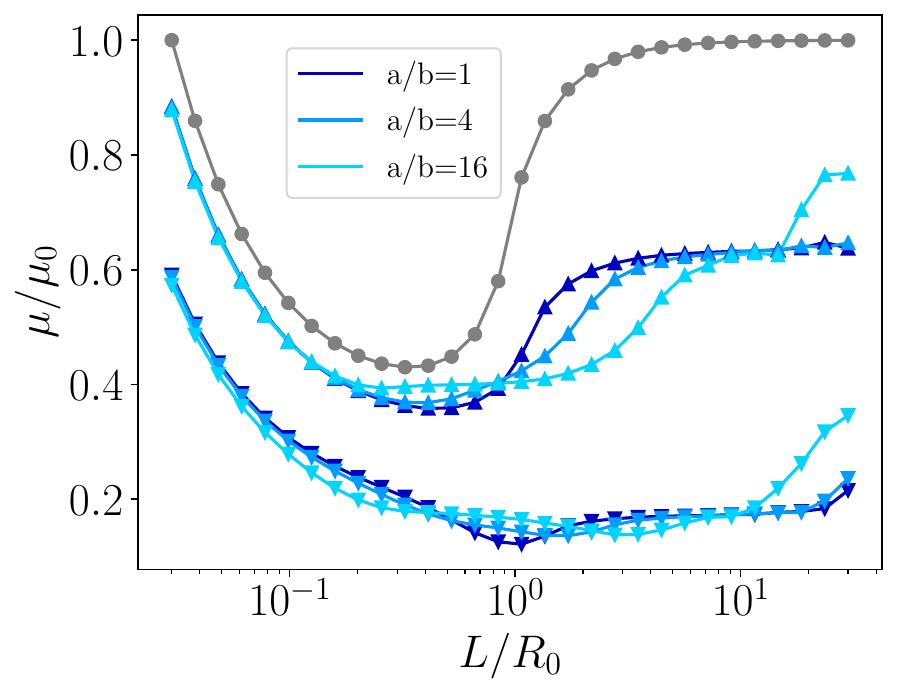}
    \caption{Normalized flux over normalized rod length $L/h_0$ for $\Delta A_{gas}=0$ (grey dots)  $\Delta A_{gas}=0.5$ (upward triangles) and for $\Delta A_{gas}=1.1$ (downward triangles) where the colors encode the aspect ratio as reported in the legend.
    \label{fig:Fluxes}}
\end{figure}

Finally, Fig.\ref{fig:Fluxes} shows the dependence of the steady state current on the rod length for different values of the corrugation. As reported for the 2D case~\cite{Malgaretti2022}, for flat channels ($\Delta A_{gas} = 0$) the
transport is determined solely by the effective diffusion coefficient. In particular, for $L \gg R_0$ the rod is almost aligned with
the axis of the channel and the effective diffusion coefficient approaches the one of the minor axis. Upon reducing $L/R_0$ the angle between the rod and the channel axis can be wider and hence both the effective diffusion coefficient, and  the mobility decrease leading to a reduction of the current. Finally, for very small aspect ratios the diffusion tensor becomes isotropic and the current increases again. Such a scenario becomes more complicated for corrugated channels, $\Delta A_{gas} \neq 0$, and for larger values of $\Delta A_{gas}$ the current monotonically decreases upon increasing $L/R_0$.

\section{Conclusion and outlook}\label{sec:Conclusion}
%%%%%%%%%%%%%%%%%%%%%%%%%%%%%%%%%%%%%%

%%%%%%%%%%%%%%%%%%%%%%%%%%%%%%%%%%%%%%%%%%%%%%%%%%%%%
 %\subsection{Conclusion}
%%%%%%%%%%%%%%%%%%%%%%%%%%%%%%%%%%%%%%%%%%%%%%%%%%%%%
 
We considered the transport of rods across corrugated channels via extending the Fick-Jacobs approximation derived for rods in Ref.\cite{Malgaretti_2020}  to $3D$. 
Within this framework, we have characterized the effective diffusion coefficient, the  mean first passage times, and the  channel permeability. 
In particular, we have focused on channels whose section breaks the axial symmetry. Interestingly, we found that the effective local free energy of the rod is quite sensitive to the aspect ratio of the channel section (see Fig.\ref{fig:DA}).
Then, we have investigated the mean first passage time (MFPT) across the full channel (and also portion of it, see Appendix~\ref{sec:MFPT-xf}). We found that the data for different aspect ratios collapse onto a master curve once the MFPT is plotted as a function of the effective free energy barrier experienced by the rod (see Fig.\ref{fig:T1}).  This allows to predict the MFPT of rods across varying-section channels relying only on the channel shape and the rod geometry. 
Finally, we have addressed the effect of external forces on the transport of rods via both the MFPT as well as by characterizing the channel permeability to different rods. Interestingly, our data show that the permeability to different rods is not sensitive to the aspect ratio (see Fig.\ref{fig:Fluxes}), despite the sensitivity shown by the effective free energy barrier (see Fig.\ref{fig:DA}).

%The findings were checked for consistencies with respect to two-dimensional results as well as analytical solutions in two and three spatial dimensions for an isotropic rod in a straight channel with and without an external force. The numerical schemes were tested for convergence in the main discretization parameter for these simplified settings. Additionally, weakly corrugated and corrugated situations were investigated both for an isotropic rod as well as elongated rods of different length. In general, the qualitative behavior of the two-dimensional situation could be recovered for all quantities such as the effective diffusion, the mean first passage time and permeability. However, the dependence on the varying parameters such as rod length became more significant for three spatial dimensions than it was observed for two spatial dimensions.

%%%%%%%%%%%%%%%%%%%%%%%%%%%%%%%%%%%%%%%%%%%%%%%%%%%%%
% \subsection{Outlook}
%%%%%%%%%%%%%%%%%%%%%%%%%%%%%%%%%%%%%%%%%%%%%%%%%%%%%
 
%The sensitivity of the permeability with respect to the free energy barrier needs a more in depth investigation to facilitate conclusions along the lines of~\cite{Malgaretti_2020}. Further improvements are also possible in accelerating the numerical schemes to allow for smaller discretization parameters for numerically unstable parameter regions, while keeping the computational times in a reasonable range. Moreover, more general channel geometries and particles with respect to shape and charge would be worth investigating.

%%%%%%%%%%%%%%%%%%%%%%%%%%%%%%%%%%%%%%
\section*{Acknowledgments}

P.M. ad J.H. acknowledge funding
by the Deutsche Forschungsgemeinschaft (DFG, German
Research Foundation) Project-ID 416229255-SFB 1411.

%%%%%%%%%%%%%%%%%%%%%%%%%%%%%%%%%%%%%%

%\begin{verbatim}
\bibliography{literature}
%\end{verbatim}

\newpage
\appendix

\section{Reduction to 2D}\label{SEC:Special_cases_3D}
%%%%%%%%%%%%%%%%%%%%%%%%%%%%%%%%%%%%%%%%%%%%%%%%%%%%%%

In the following, we identify special cases and, in particular, discuss the consistancy with the two-dimensional setting outlined in~\cite{Malgaretti_2020}.

\begin{itemize}
\item 
If we set $\Phi=0$ and $y_M=0$, we recover the two-dimensional situation since with $\sin(\Phi)=\sin(0)=0$ and $b=h$ it holds
\begin{align}
    z_M + L\cos(\theta)=h.
\end{align}
However, we remark that $\theta\neq \theta_{2D}$, but $\theta=\frac{\pi}{2}-\theta_{2D}$ and consequently one has\\  $\cos(\theta)=\cos(\frac{\pi}{2}-\theta_{2D})=\sin(\theta_{2D})$, which leads to the following expression as derived in~\cite{Malgaretti_2020}:
\begin{align}
    \theta_m=
    \begin{cases}
    \frac{\pi}{2} - \arccos\left(\frac{h(x)-|z_M|}{L}\right) & h(x)-L\leq|z_M|\leq h(x)\\
    \frac{\pi}{2} & |z_M|<h(x)-L
    \end{cases}\\
    \theta_M=
    \begin{cases}
    \frac{\pi}{2} + \arccos\left(\frac{h(x)-|z_M<}{L}\right) & h(x)-L\leq|z_M|\leq h(x)\\
    \frac{\pi}{2} & |z_M|<h(x)-L.
    \end{cases}
\end{align}
\item 
Likewise, if we consider a large aspect ratio for the channel with elliptic cross section, i.e. $\frac{a}{b}\rightarrow\infty \Leftrightarrow \frac{b}{a}\rightarrow 0$, it holds
\begin{align}
    \lim_{\frac{b}{a}\rightarrow 0} \left( (y_M+L\sin(\theta)\sin(\Phi))^2\frac{b^2}{a^2} + (z_M+L\cos(\theta))^2\right) = b^2,\\
    z_M + L\cos(\theta) =b=h
\end{align}
i.e. we again recover the two-dimensional situation.
\end{itemize}

\section{MFPT as a function of the final position $x_f$}\label{sec:MFPT-xf}

\begin{figure}%[h]
    \centering
    \includegraphics[width=0.4\linewidth]{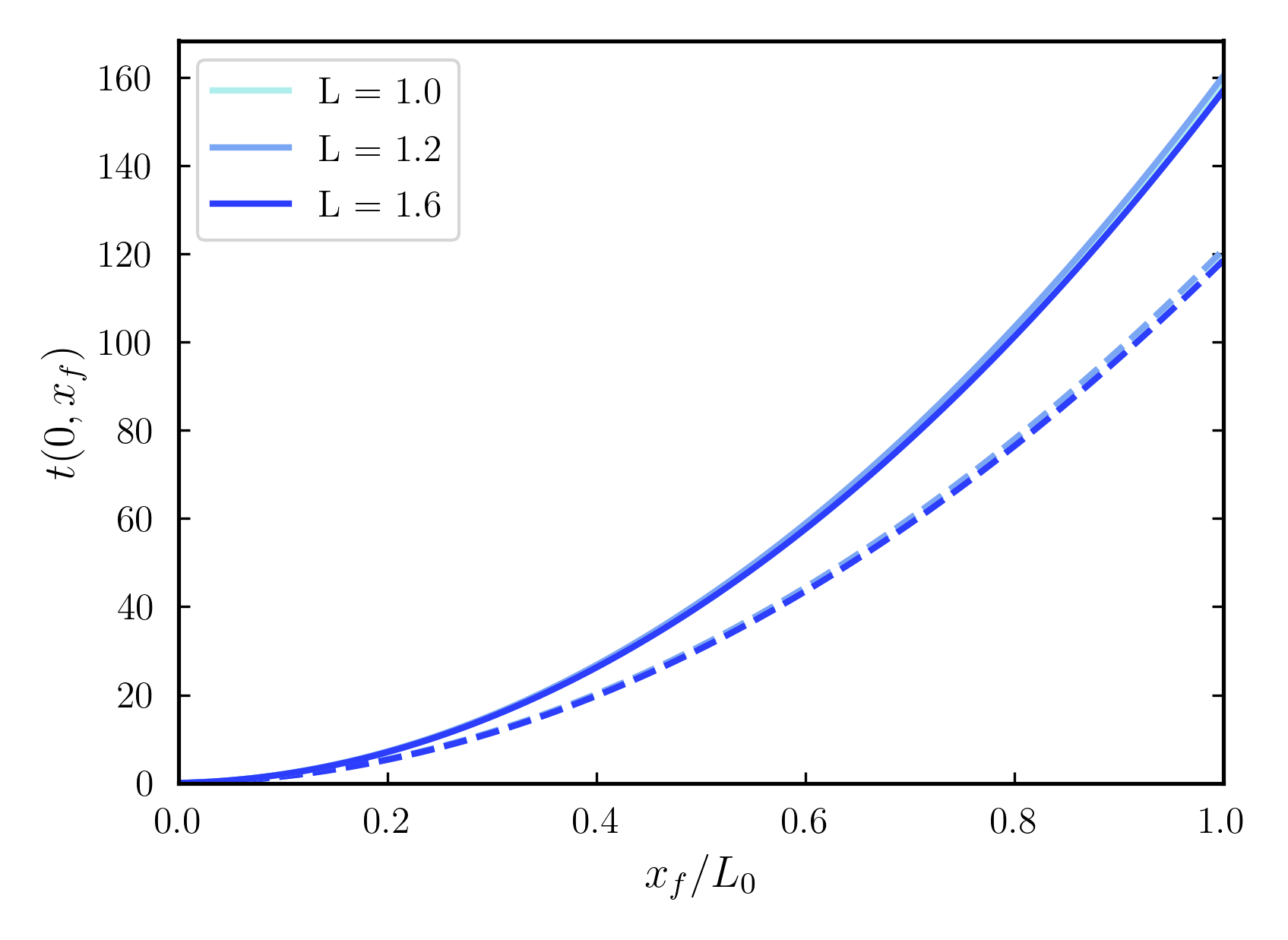}
    \includegraphics[width=0.4\linewidth]{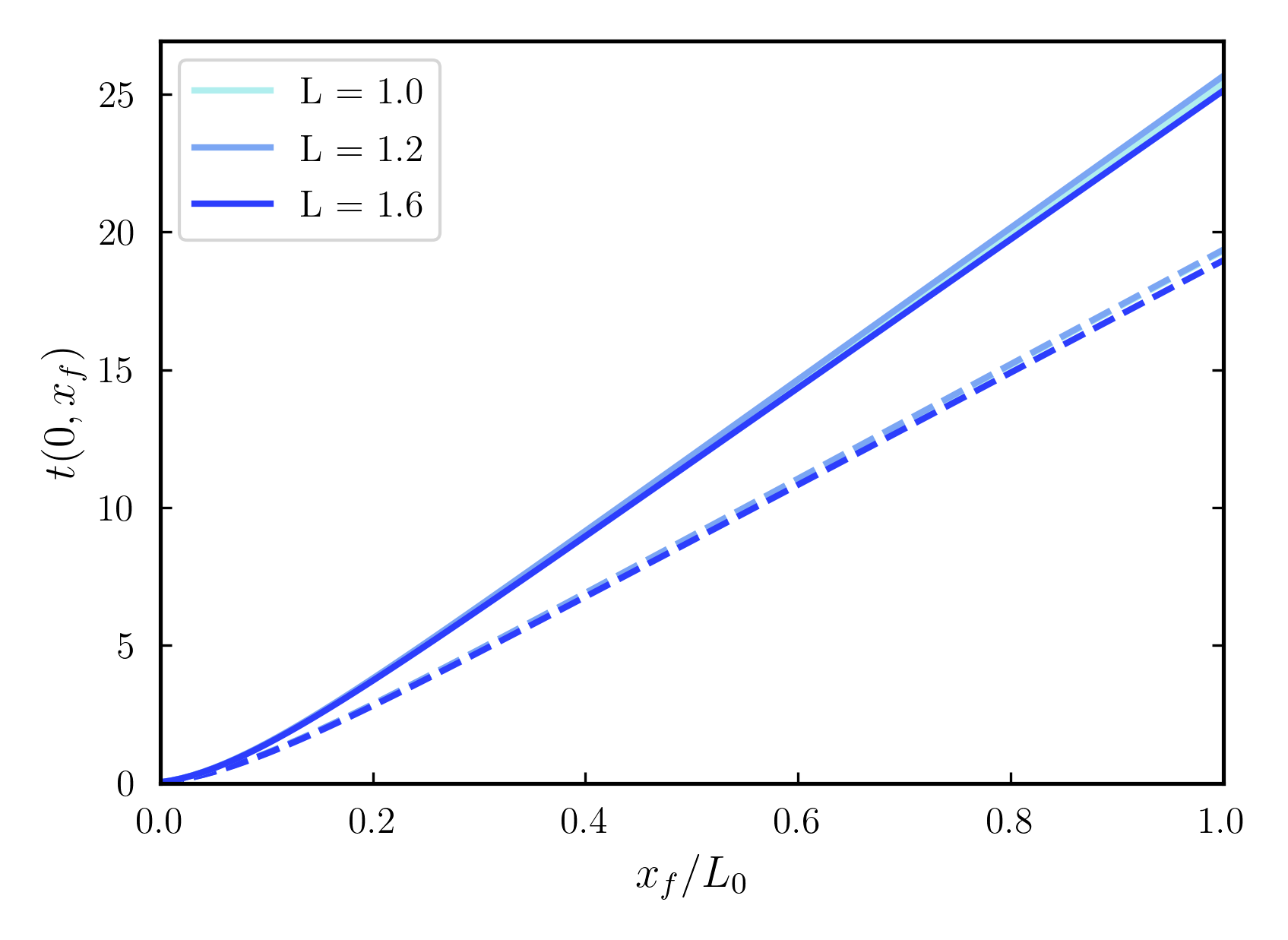}
    \includegraphics[width=0.4\linewidth]{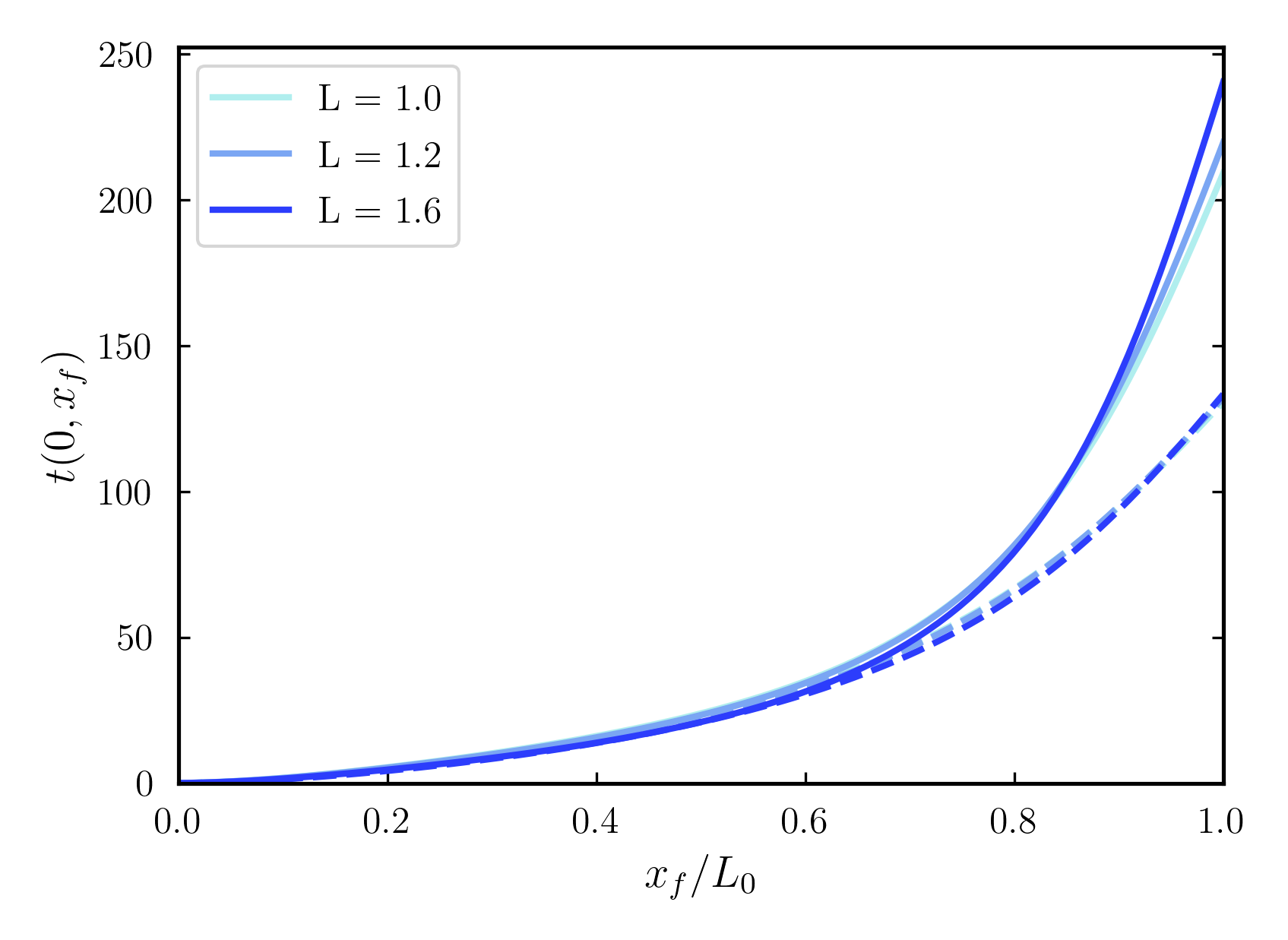}
    \includegraphics[width=0.4\linewidth]{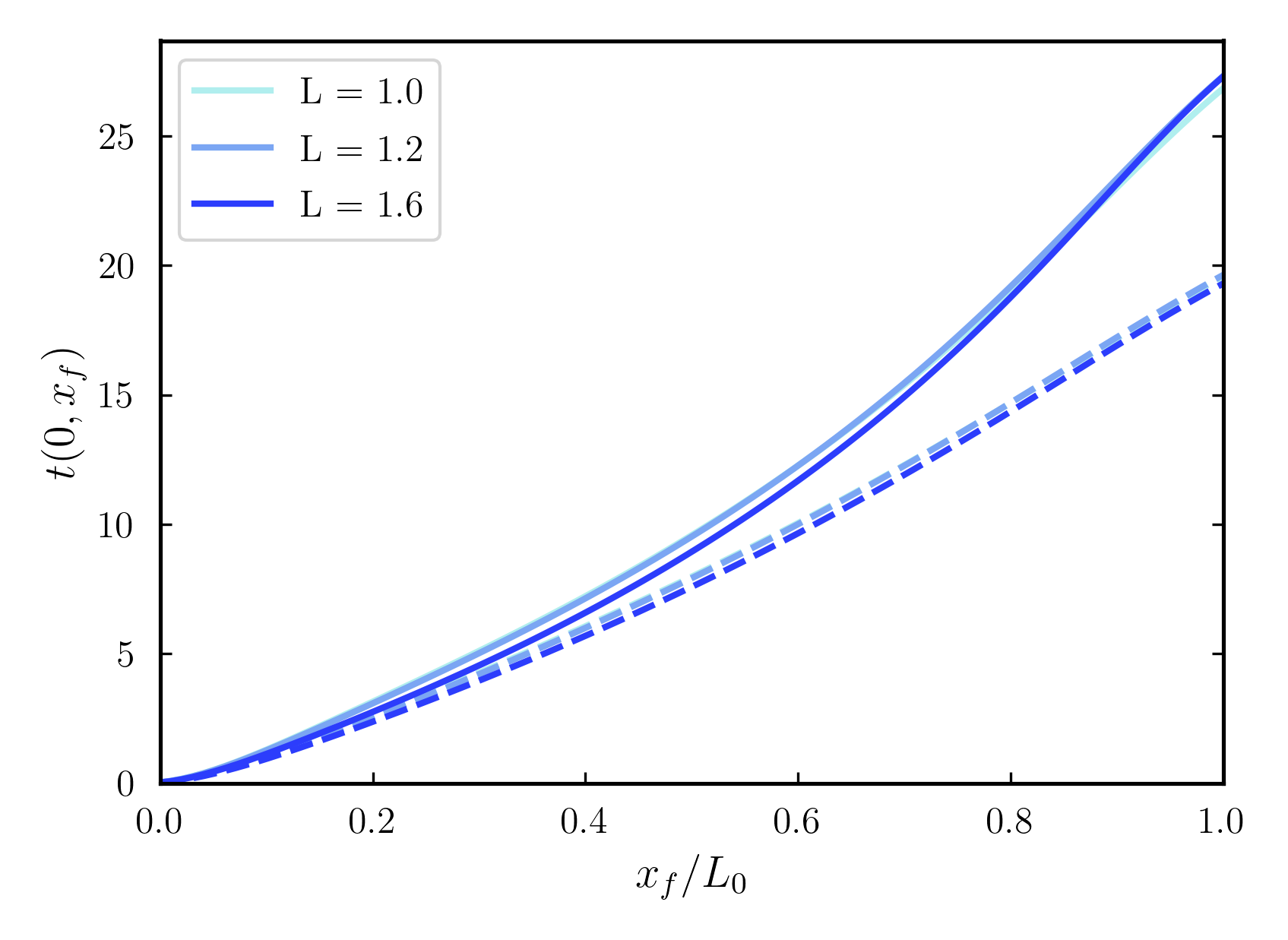}
    \includegraphics[width=0.4\linewidth]{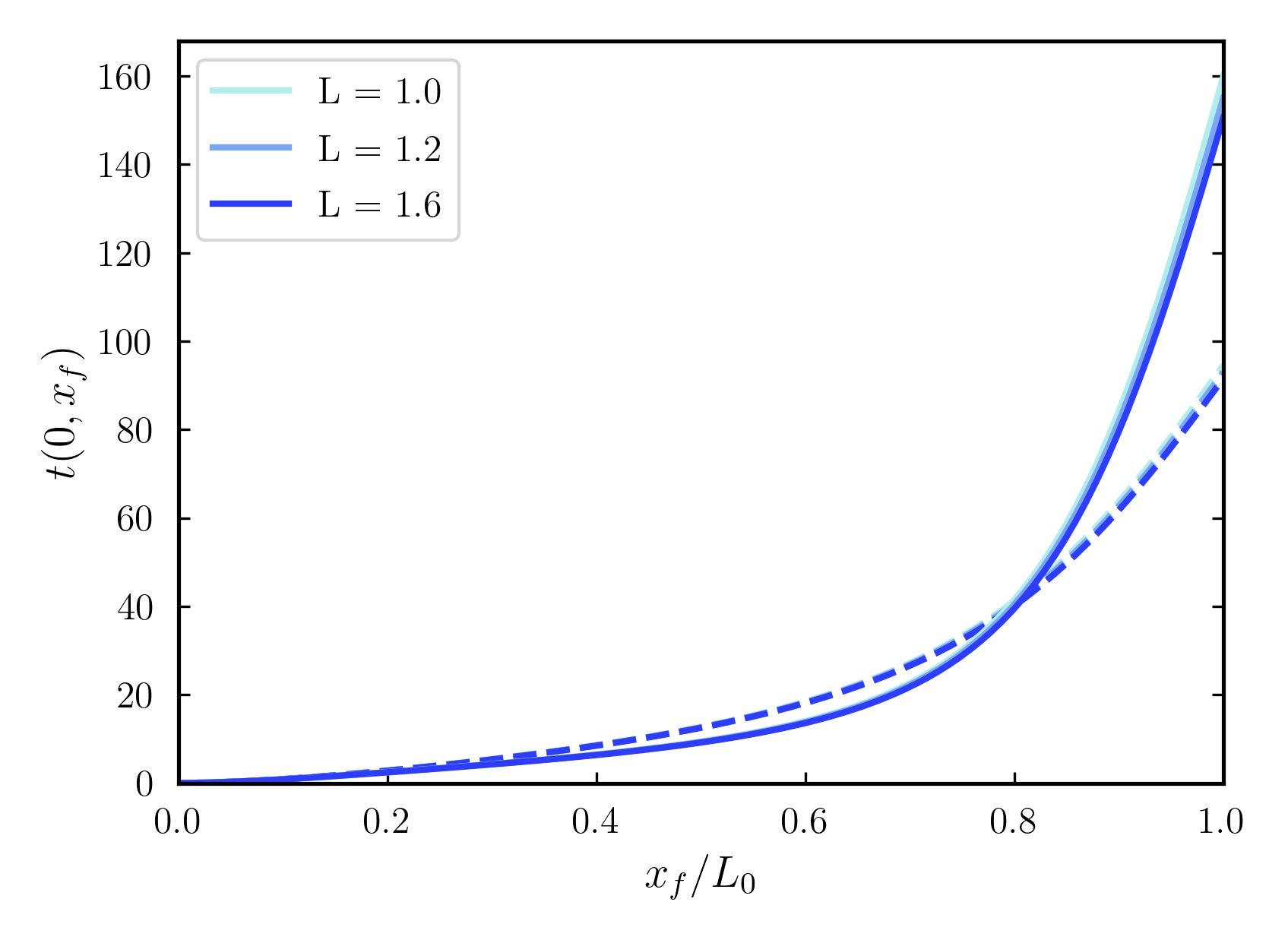}
    \includegraphics[width=0.4\linewidth]{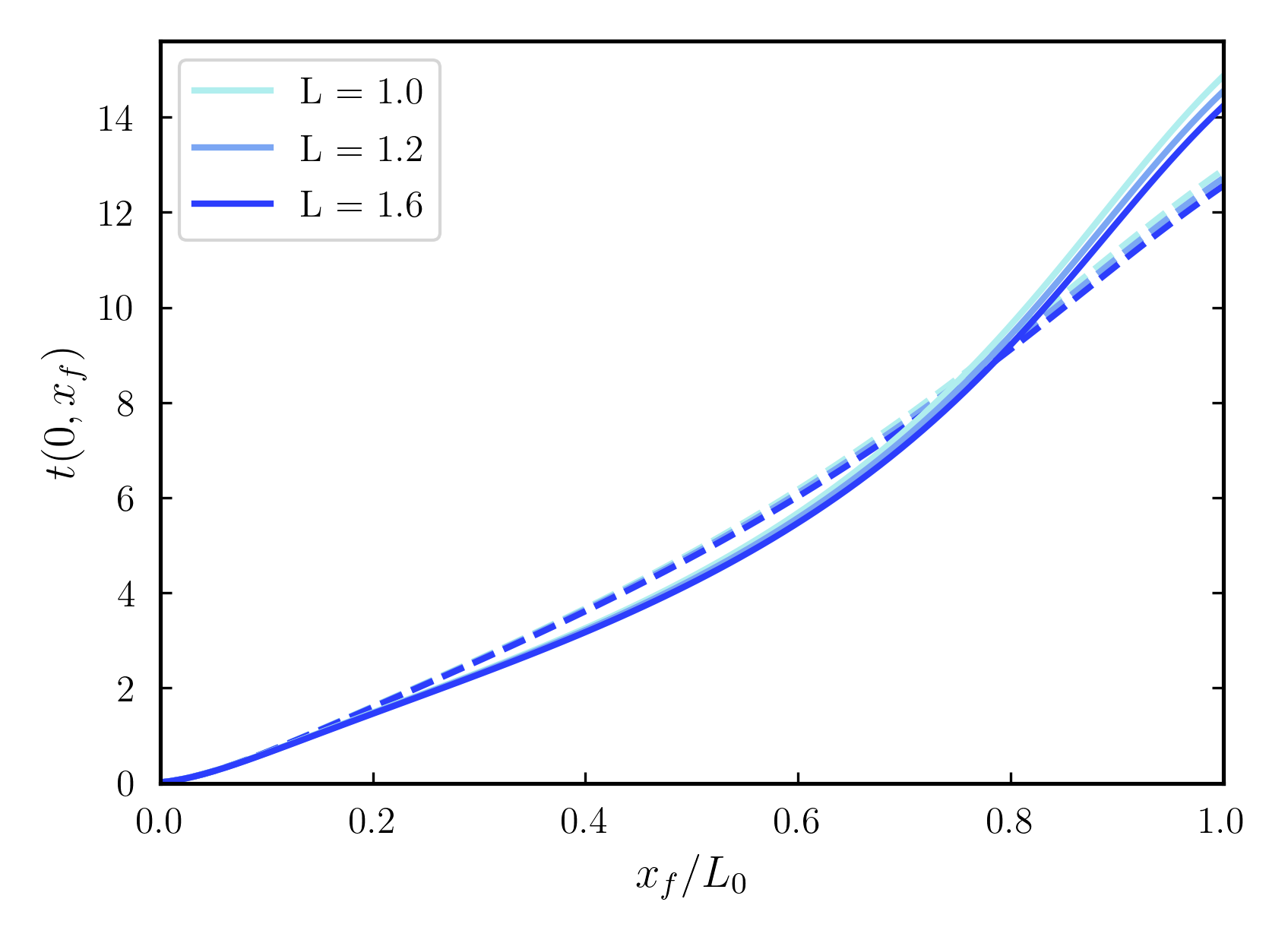}
    \caption{Mean first passage time $t(0,x_f)$ in 2D (dashed lines) and 3D (solid lines) until normalized position $x_f/L_0$ is reached starting from the origin $x=0$ in a straight (top row, 2D: $h_0=3 \mu m, h_1=0 \mu m$, 3D: $R_0=3 \mu m, R_1=0 \mu m$), weakly corrugated (middle row, 2D: $h_0=3 \mu m, h_1=0.3 \mu m$, 3D: $R_0=3 \mu m, R_1=0.3 \mu m$), and strongly corrugated (bottom row, 2D: $h_0=0.5 \mu m, h_1=0.3 \mu m$, $R_0=0.5 \mu m, R_1=0.3 \mu m$) channel. Thin rod of radius $l=0.15 \mu m$ without external force, i.e. $f=0$, (left column) and with external force, i.e. $f=1$, (right column) for varying rod lengths $L=1 \mu m$ (light blue) , $L=1.2 \mu m$ (blue), and $L=1.6 \mu m$ (dark blue). }
    \label{fig:Validation_time_sphere}
\end{figure}

We investigate the mean first passage time of a thin rod through a straight, weakly and strongly corrugated channel without external force and for a homogeneous force. It is evident from the results depicted in Figure~\ref{fig:Validation_time_sphere} that for, $x_f<L/2$, \pa{the transport is mainly controlled by the effective entropic force which is alway pulling the positive direction, and hence the time grows almost linearly with $x_f$. In contrast, for $x_f>L/2$ the entropic force is pushing along the opposite direction. This implies that the rod has to diffuse against the entropic force, leading to a steeper increase of $t_1$ on $x_f$. }
We remark that the rod is starting in the most corrugated part of the channel which is thereafter widening up.
%the the transport is mainly the limiting linear and quadratic behaviors as discussed in Section~\ref{SEC:Limits_MFPT}\NR{This does not exist anymore!} are recovered. \NR{I suggest to explicitly outline parts of the commented text from the previous appendix and put it in context to the referenced section on MFPT. Not completely clear to me how detailed to do that best as MFPT is wrt to $A$ whereas here we need $f$: In the limit of vanishing forces or small travel distances $\beta f L \ll 1$ we get quadratic growth
%\begin{align}
%    t(0) \simeq \frac{L^2}{D}
%\end{align}
%whereas for large $L$ we obtain a linear relationship, i.e. $t(0) \simeq L$.} 
We further observe that the rod generally travels slower through the three-dimensional channel opposed to the two-dimensional channel as it has more degrees of freedom of movement and can be distracted from the streamlined flow through the channel center.

%\NR{Wrt to rod length I do not see a clear trend and the intersection of 3D and 2D curves for strongly corrugated channel might be due to numerics or physics??.)}%end NR

\newpage

\begin{figure}[h]
    \centering
    \includegraphics[width=0.4\linewidth]{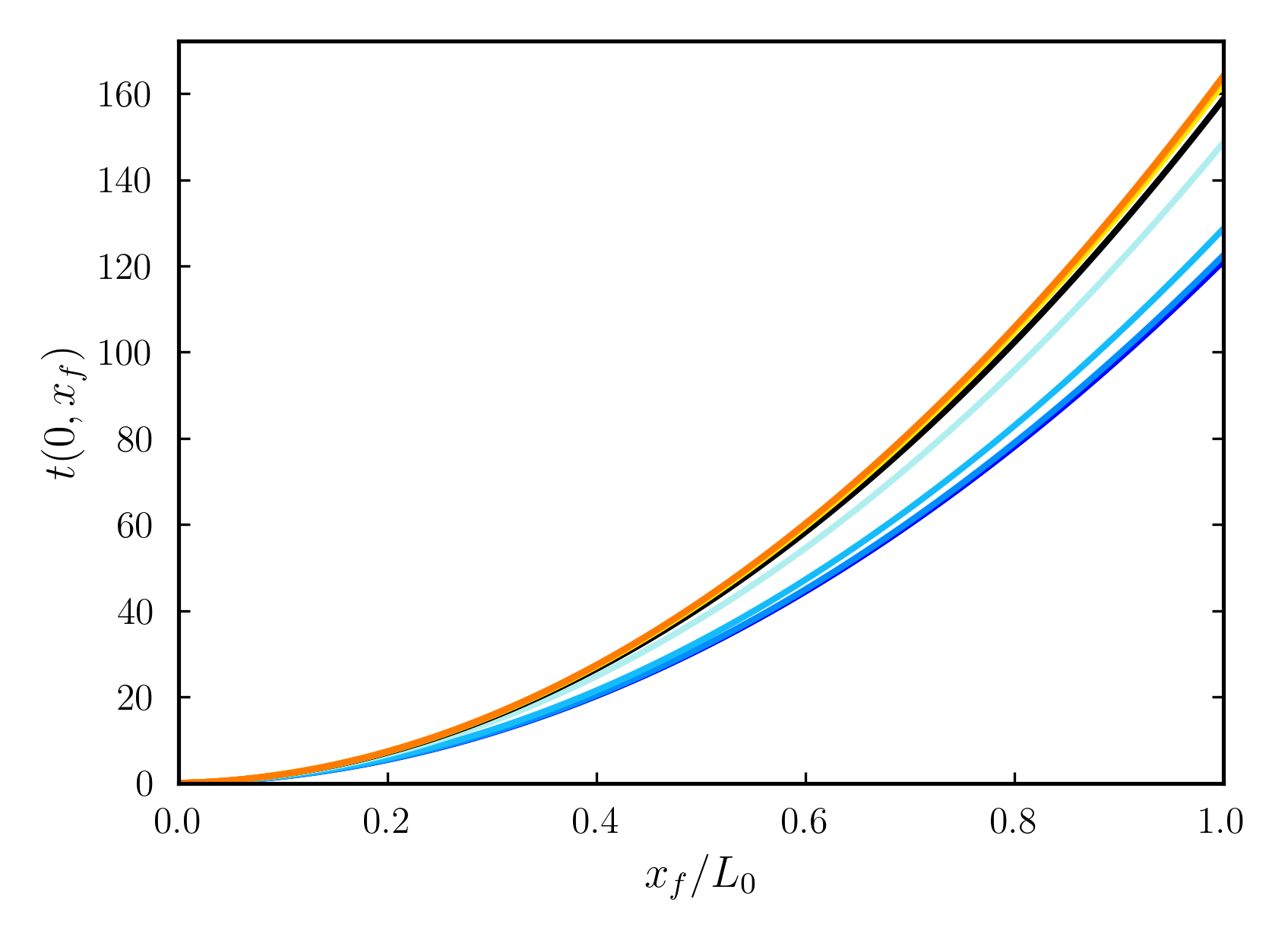}
    \includegraphics[width=0.4\linewidth]{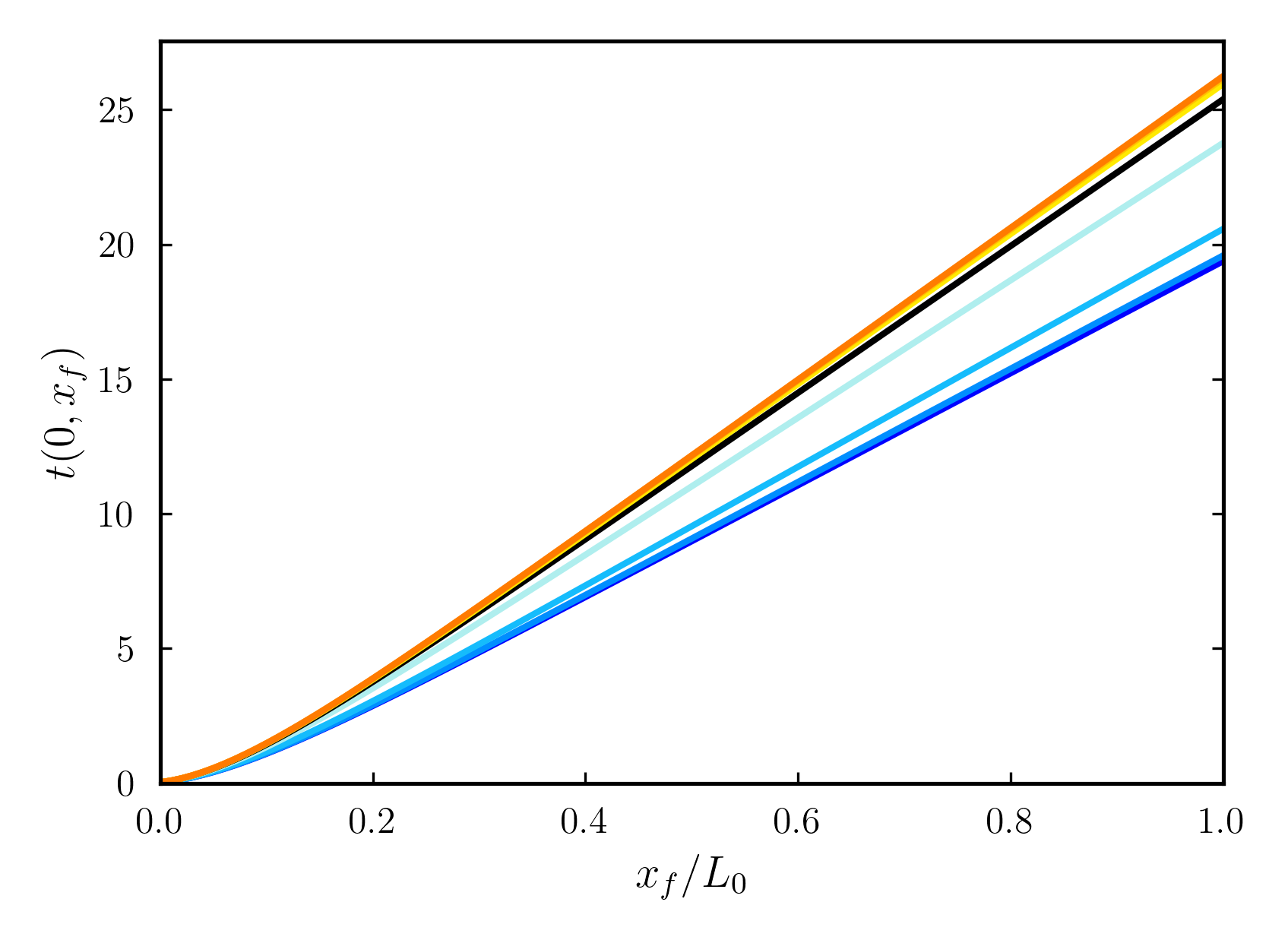}\\
    \includegraphics[width=0.4\linewidth]{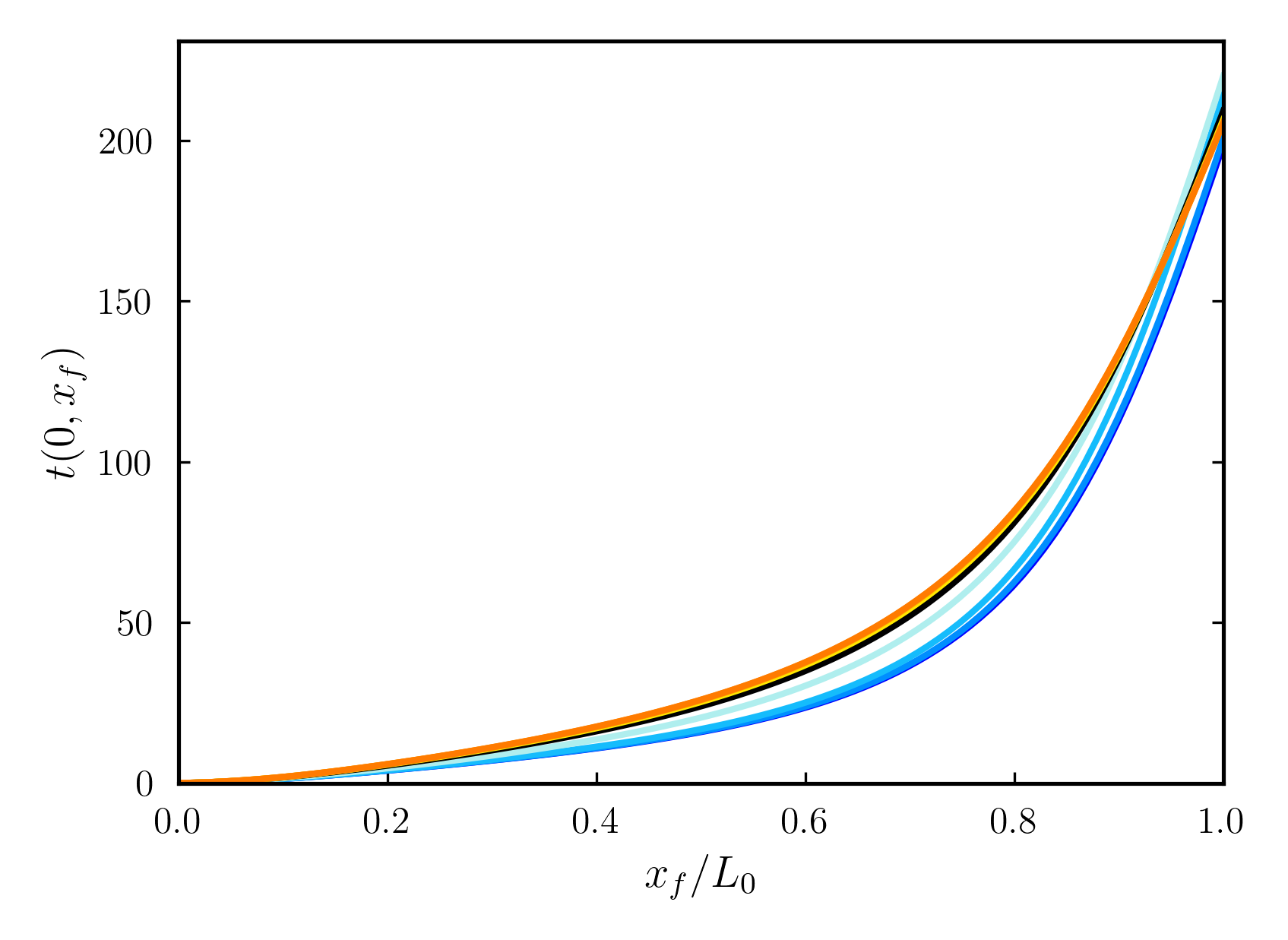}
    \includegraphics[width=0.4\linewidth]{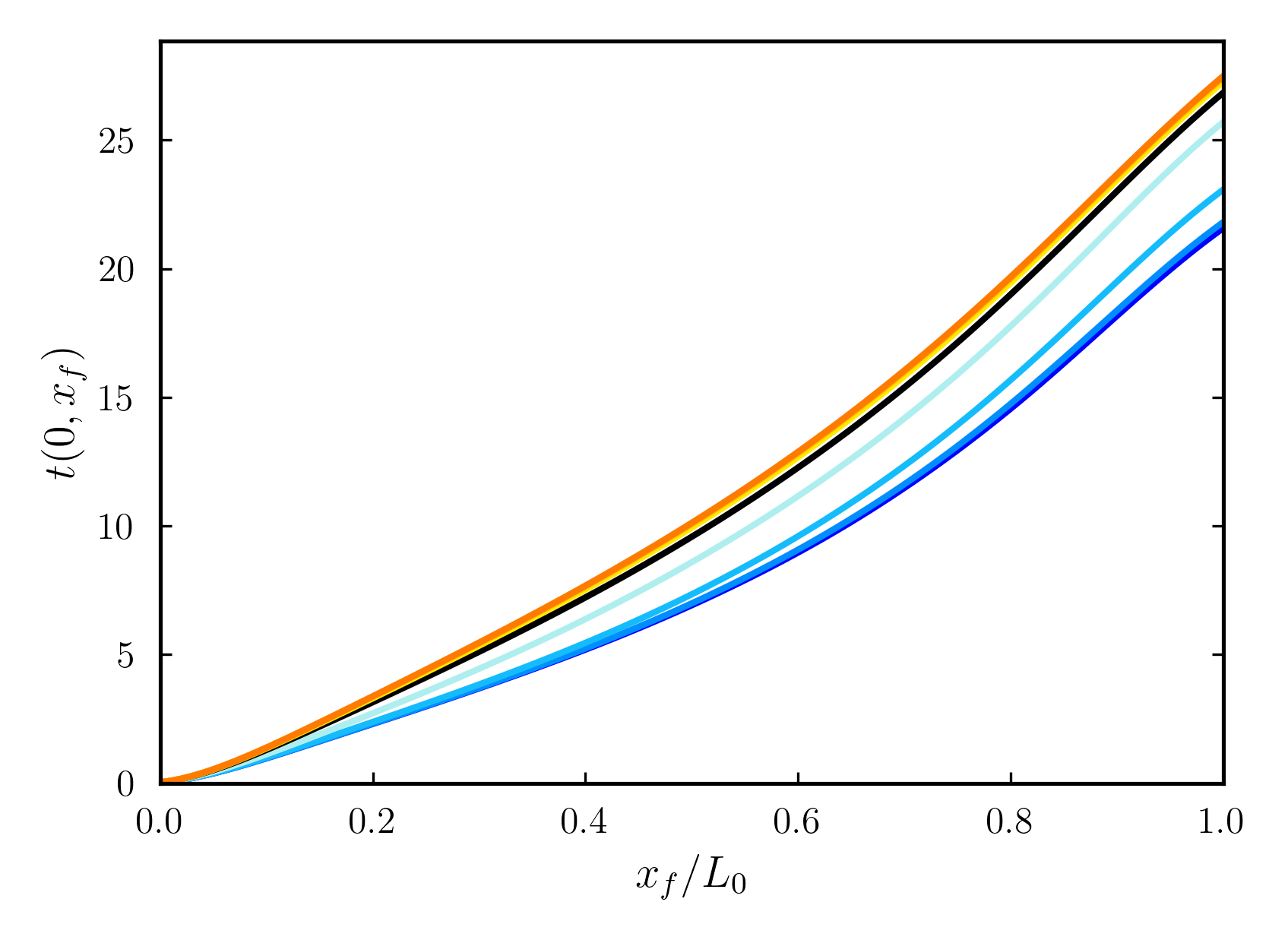}\\
    \includegraphics[width=0.4\linewidth]{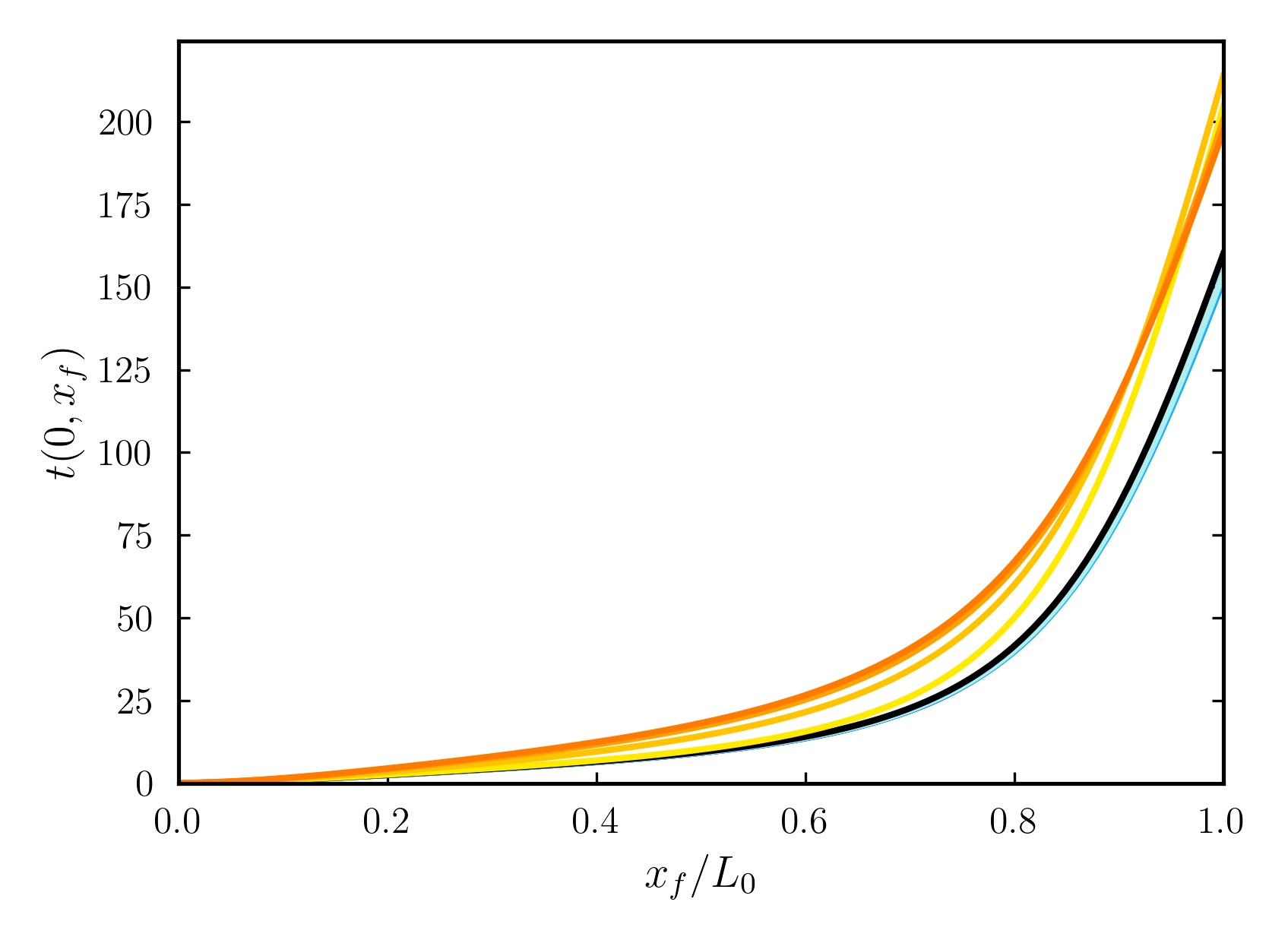}
    \includegraphics[width=0.4\linewidth]{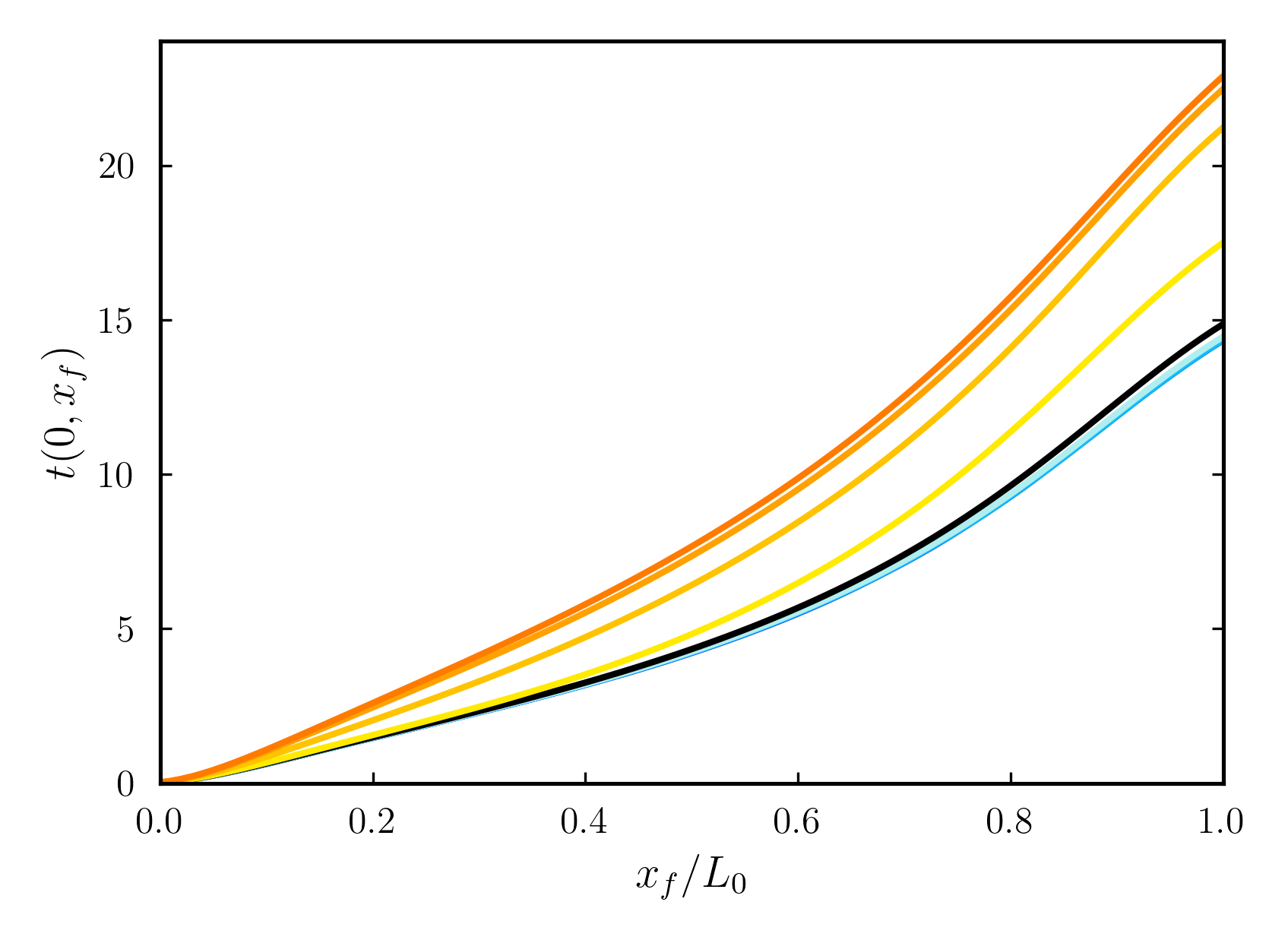}
\caption{
Mean first passage time $t(0,x_f)$ until normalized position $x_f/L_0$ is reached starting from the origin $x=0$ in a straight (top row, $R_0=3 \mu m, a=b=0 \mu m$), weakly corrugated (middle row, $R_0=3 \mu m, a=b=0.3 \mu m$), and strongly corrugated (bottom row, $R_0=0.5 \mu m, a=b=0.3 \mu m$) three-dimensional channel with (stretched) elliptic cross section with aspect ratios $\lambda$ using scaling factors $\lambda= [0.0625, 0.125, 0.25, 0.5, 1, 2, 4, 8, 16]$, and main elliptic axes $R_0$ and $R_0\lambda$. Circular cross section (black line).     
Thin rod of radius $l=0.15 \mu m$ and length $L=1 \mu m$ without external force, i.e. $f=0$, (left column) and with external force, i.e. $f=1$ (right column). 
        }
    \label{fig:Time_elliptic_f1-big}
\end{figure}

It is evident that for very stretched ellipses with large aspect ratios (dark blue lines), the MFPT approaches the MFPT of the two-dimensional setting for the straight channel, compare also Figure~\ref{fig:Validation_time_sphere}, whereas a further widening of the circular cross section does not lead to a large increase in the MFPT as already for the circular cross section, the rods have a high degree of freedom in their movement. 
For the corrugated channels, the results in Figure~\ref{fig:Time_elliptic_f1-big} must be interpreted also in the context of Figure~\ref{fig:DA}. In particular for the strongly corrugated channel, we have $L/R_0=2$ and the maximum of the free energy barrier has not been (completely) passed for large aspect ratios. 

%\NR{How to explain that more detailed?}

\newpage

Interestingly black line indicating the MFPT for the circular cross section is slowest for both straight and weakly corrugated channel, but fastest for strongly corrugated channel. Again this can be interpreted in the context of Figure~\ref{fig:DA} as for the strongly corrugated channel, we have $L/R_0=2$, but the maximum of the free energy barrier has maybe not been completely passed. Along the line of Figure~\ref{fig:DA}, the results confirm that area as well as shape play an important role for transport.

\begin{figure}[h]
    \centering
    \includegraphics[width=0.4\linewidth]{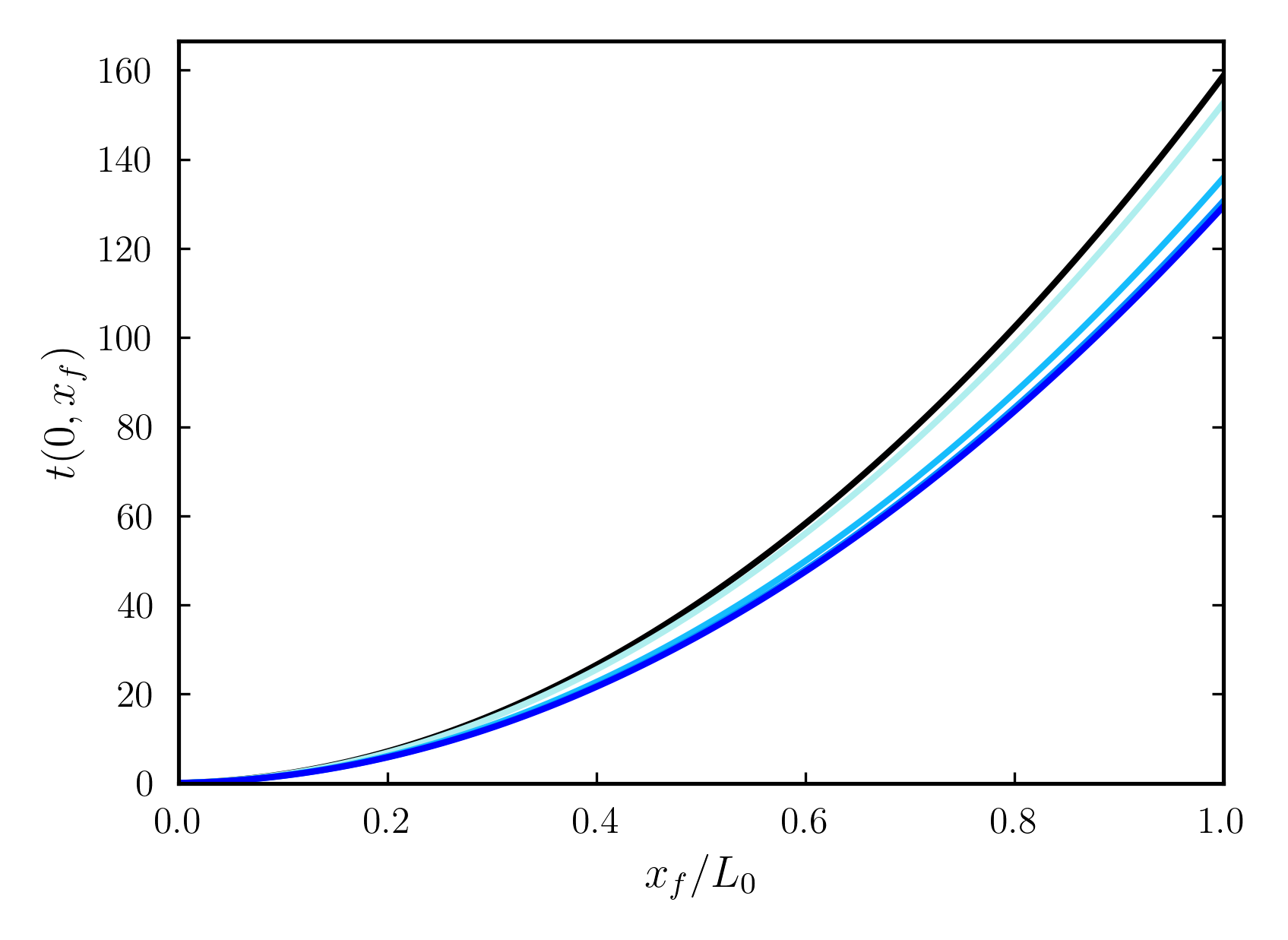}
    \includegraphics[width=0.4\linewidth]{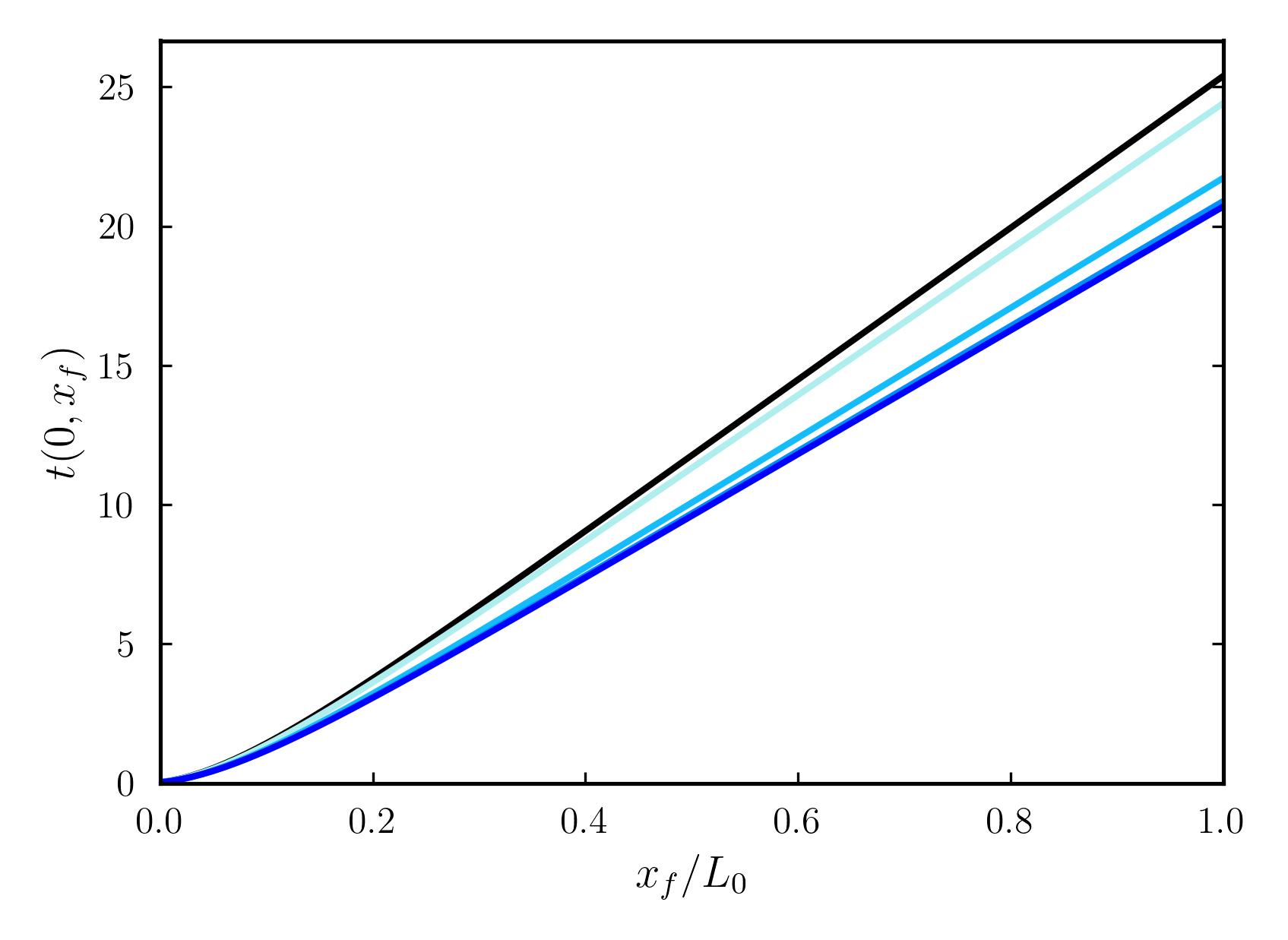}\\
    \includegraphics[width=0.4\linewidth]{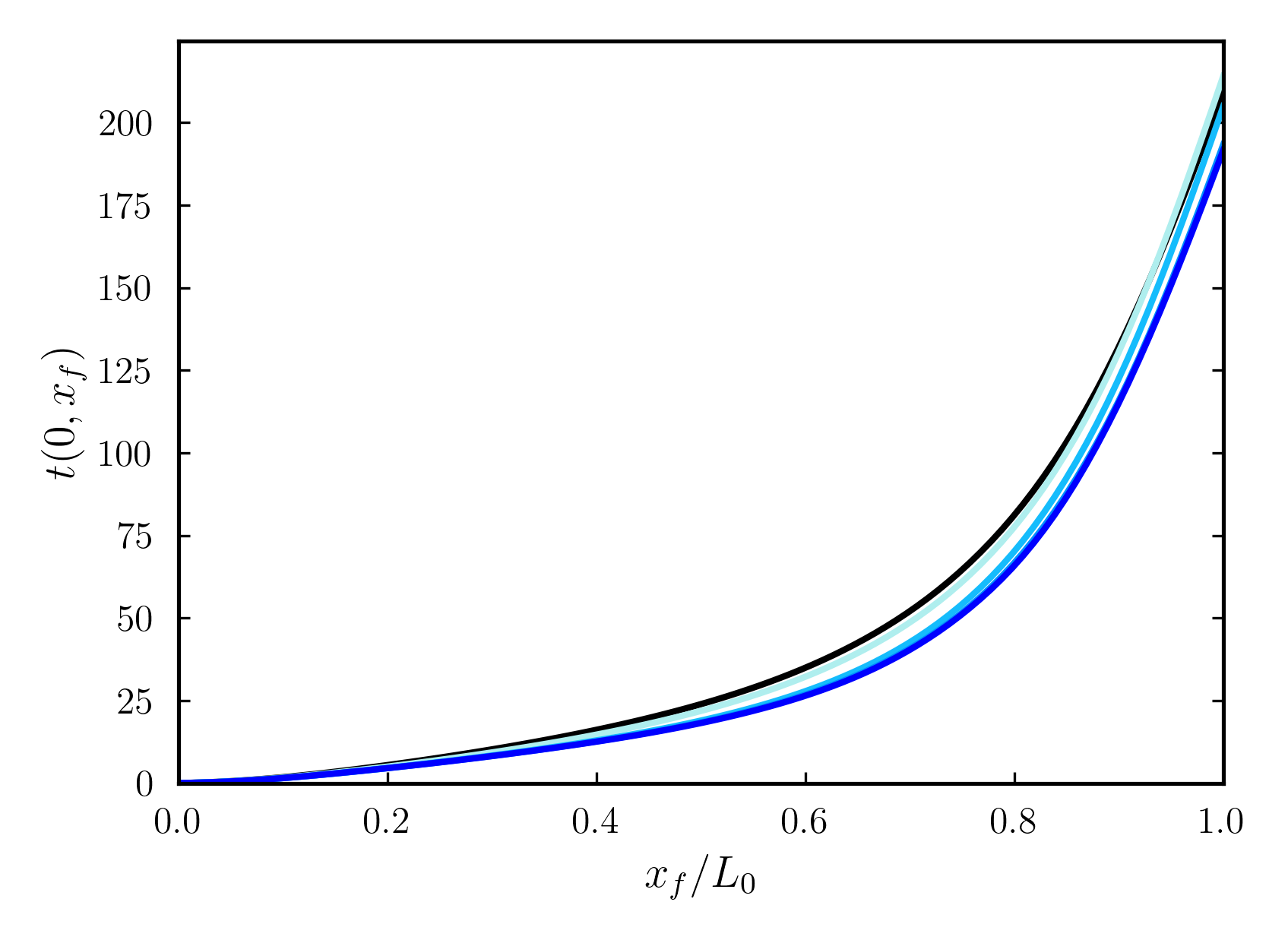}
    \includegraphics[width=0.4\linewidth]{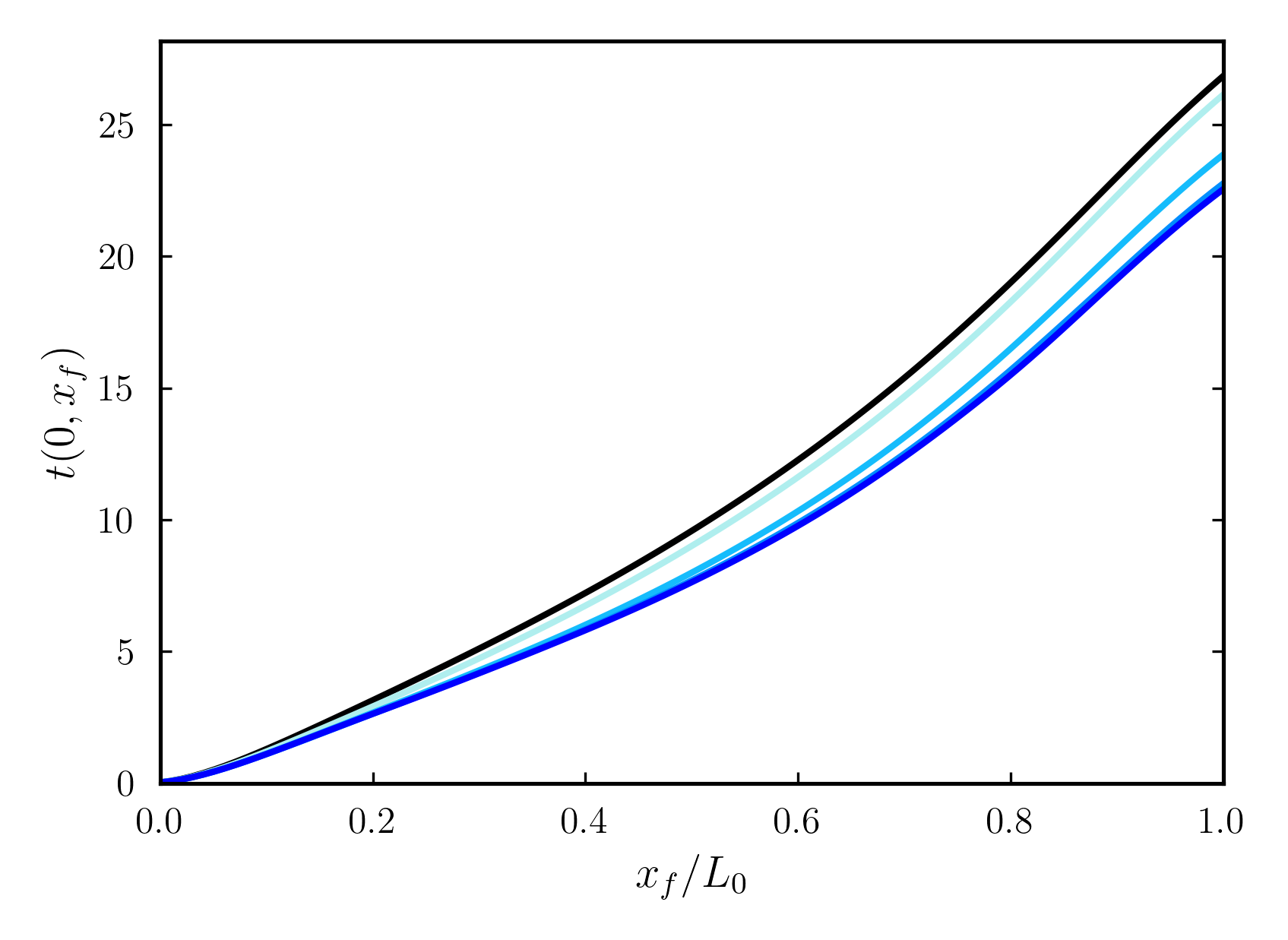}\\
    \includegraphics[width=0.4\linewidth]{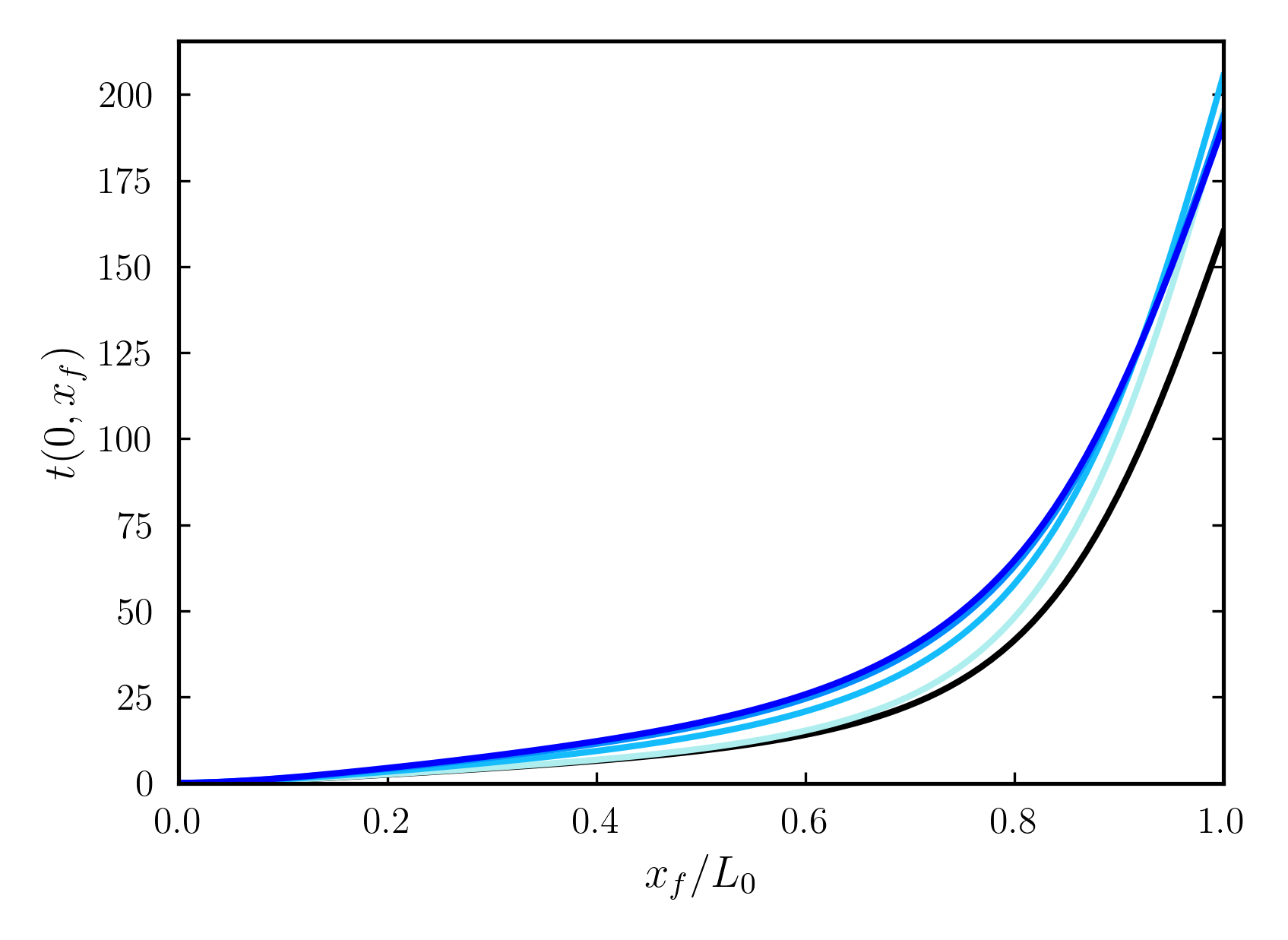}
    \includegraphics[width=0.4\linewidth]{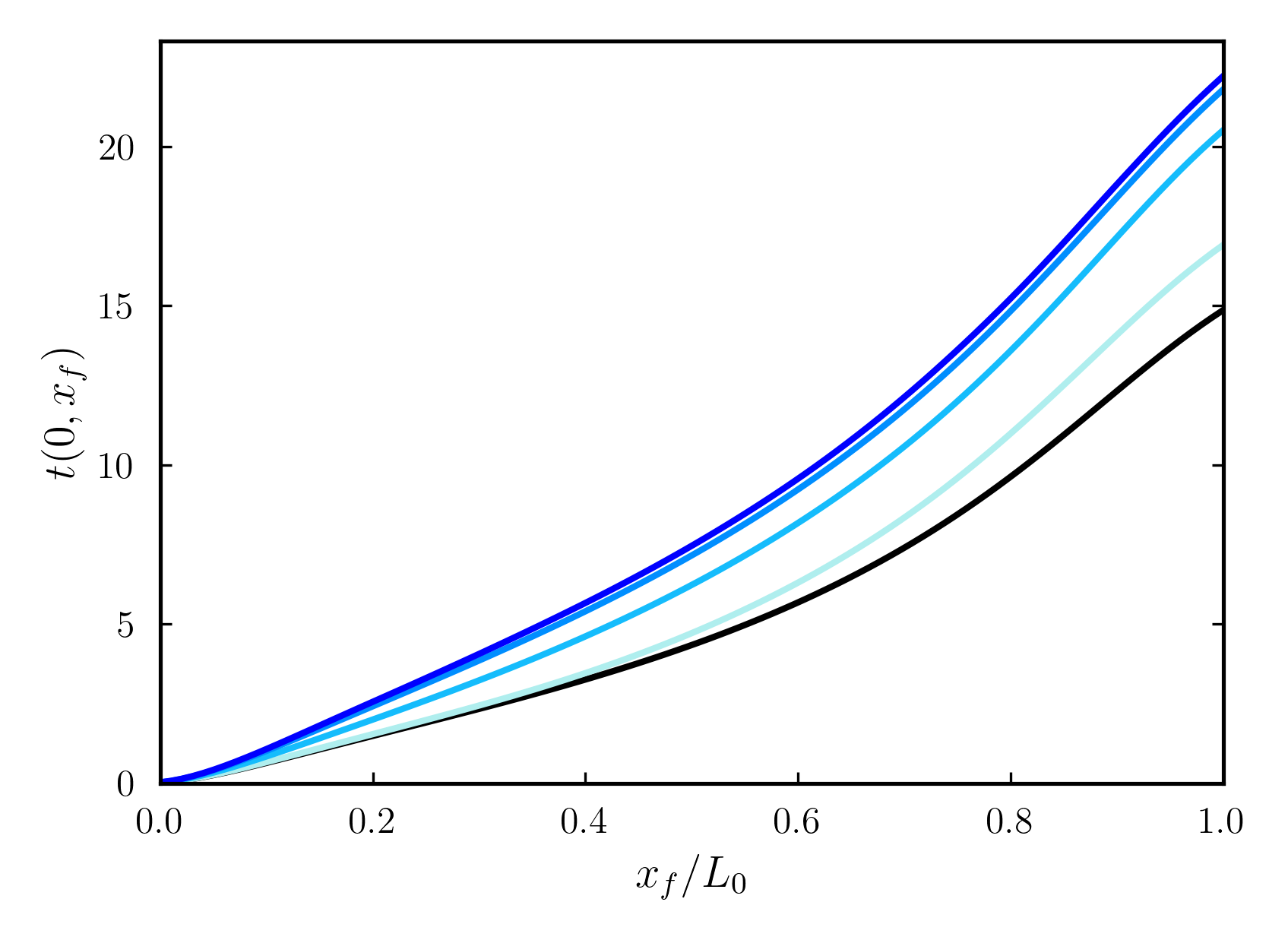}
\caption{Mean first passage time $t(0,x_f)$ until normalized position $x_f/L_0$ is reached starting from the origin $x=0$ in a straight (top row, $R_0=3 \mu m, a=b=0 \mu m$), weakly corrugated (middle row, $R_0=3 \mu m, a=b=0.3 \mu m$), and strongly corrugated (bottom row, $R_0=0.5 \mu m, a=b=0.3 \mu m$) three-dimensional channel with elliptic cross section with constant area, but different shape using scaling factors $\lambda= 1, 2, 4, 8, 16] $, and main elliptic axes $R_0/\lambda$ and $R_0\lambda$. Circular cross section (black line).     
Thin rod of radius $l=0.15 \mu m$ and length $L=1 \mu m$ without external force, i.e. $f=0$, (left column) and with external force, i.e. $f=1$, (right column). }
    \label{fig:Time_elliptic_f1-big-app}
\end{figure}

\end{document}